\documentclass{article}
\usepackage{amsfonts}
\usepackage{amsmath}
\usepackage{amssymb}
\usepackage{color}
\usepackage{eurosym}

\newtheorem{theorem}{Theorem}

\newtheorem{example}{Example}

\begin{document}

\title{Revenue sharing at music streaming platforms\thanks{%
Financial support from Ministerio de Ciencia e Innovaci\'{o}n, MCIN/AEI/
10.13039/501100011033, through grants PID2020-113440GBI00 and
PID2020-115011GB-I00, Junta de Andaluc\'{\i}a through grant P18-FR-2933, and
Xunta de Galicia through grant ED431B2022/03 is gratefully acknowledged. We
thank audiences at Paris, Istanbul and Oviedo for helpful comments and
suggestions.}}
\author{\textbf{Gustavo Berganti\~{n}os}\thanks{%
ECOBAS, Universidade de Vigo, ECOSOT, 36310 Vigo, Espa\~{n}a} \\
\textbf{Juan D. Moreno-Ternero}\thanks{%
Department of Economics, Universidad Pablo de Olavide, 41013 Sevilla, Espa%
\~{n}a; jdmoreno@upo.es}}
\maketitle

\begin{abstract}
We study the problem of sharing the revenues raised from subscriptions to
music streaming platforms among content providers. We provide direct,
axiomatic and game-theoretical foundations for two focal (and somewhat
polar) methods widely used in practice: \textit{pro-rata} and \textit{%
user-centric}. The former rewards artists proportionally to their number of
total streams. With the latter, 
each user's subscription fee is proportionally divided among the artists
streamed by that user. 
We also provide foundations for a family of methods compromising among the
previous two, which addresses the rising concern in the music industry to
explore new streaming models that better align the interests of artists,
fans and streaming services.
\end{abstract}

\bigskip

\medskip {} 


\bigskip

\bigskip

\newpage

\bigskip

\bigskip

\section{Introduction}

Steve Jobs unveiled the iPod on October 23, 2001. As he put it himself a few
years later, that event \textit{didn't just change the way we all listen to
music, it changed the entire music industry}. And it is difficult to deny
such a statement. Because digital music had started to emerge at the end of
the 20th century, mostly thanks to the advent of peer-to-peer (P2P)
technologies. But successful file-sharing platforms, such as Napster, were
under scrutiny by the music industry, which aggressively pursued stronger
copyright enforcement and regulations (e.g., Bhattacharjee et al., 2007). In
that environment, Apple managed to persuade record companies to sale
individual tracks for 99 cents. Gradually, the industry found a new way to
stay profitable and even embraced new technology advances like streaming. A
decade later, Spotify, an early forerunner music streaming platform,
announced a customer base of 1 million paying subscribers across Europe, and
was officially launched in the US. Alternative services (such as Beats
Music, Amazon Music Unlimited and Google Play Music All-Access, as well as a
new Apple Music service) also emerged in the market. According to Statista,
revenue in the Music Streaming market is projected to reach US$\$25.84$
billions in 2023. The number of users is expected to amount to $1.1$ billion
by 2027.%
\footnote{%
https://www.statista.com/outlook/dmo/digital-media/digital-music/music-streaming/worldwide 
} 

Streaming platforms generate massive amounts of revenue nowadays. Aleei et
al. (2022) argue that \textquotedblleft streaming services generated $\$4.3$
billion in the first half of 2019 (with $77\%$ of that coming from paid
subscriptions)\textquotedblright . Users typically pay a fixed (monthly)
amount to freely access their libraries. 
A common practice for platforms is to distribute around $70\%$ of the
revenue received from subscriptions among artists (e.g., Meyn et al., 2023).
Platforms also raise money from other sources (for instance, advertisements)
but the most important source are subscriptions and we shall concentrate on
them here. An ensuing interesting problem is to allocate the corresponding
part of those revenues among participating artists, based on their streaming
times. This will be the object of study in this paper.


We introduce a stylized model, whose ingredients are three: a group of
artists, a group of users and a matrix indicating the streaming times each
user played each artist. Based on this input, a popularity \textbf{index},
which measures the importance of each artist, is constructed. The reward
received by each artist from the revenues generated in each problem is based
on such a popularity index. The most frequent indices are the so called 
\textbf{pro-rata} index, which renders artists rewarded proportionally to
the total number of streams and the so called \textbf{user-centric} index,
which renders artists rewarded so that the revenue generated by each user is
divided among the artists listened by the user proportionally to the total
number of streams.\footnote{%
Typically, a user pays around $\$10$ per month. As 70\% of the amount
generated by users is devoted to pay artists, we could just consider in our
model each user pays $\$7$. For ease of exposition, we shall assume that
each user pays a normalized amount of $1$ to have unlimited access to the
contents in the streaming platform, during the period of time (usually, a
month). Note also that, in practice, it is considered that an artist
obtained a streaming unit when a user played a song from that artist for at
least 30 seconds (e.g., Meyn et al., 2023). Thus, the entry of the
(streaming) matrix $t_{ij}$ would be the number of streaming units artist $i$
obtained from user $j$ during one month. There is, nevertheless, a debate in
the industry about how streams should be computed. Another alternative is a
remuneration based on per-second usage (e.g., Meyn et al., 2023). Our
theoretical model is general and allows us to accommodate other ways of
measuring streams.} Then, the amount received by each artist is computed by
adding the amounts obtained from each user.

We take several approaches to analyze our model. In the first (axiomatic)
approach, we present axioms that formalize normatively appealing principles
for popularity indices. Some convey structural ideas that reflect
operational features of the index. For instance, \textit{additivity} says
that if we can present a problem as the sum of smaller problems (such as in,
say, multinational platforms), then the index in the original problem should
coincide with the sum of the indices in the smaller problems. And \textit{%
homogeneity} says that the index should reflect accordingly the cases in
which each user has reproduced content from a given artist a certain times
more than content from another artist. Other axioms (\textit{equal
individual impact of similar users} and \textit{equal global impact of users}%
) model alternative forms of \textit{marginalism}, i.e., the impact of extra
users in the platform. And we also consider axioms reflecting concerns for
fairmess: \textit{reasonable lower bounds} and \textit{click-fraud-proofness}%
. The first one states that artists should at least receive the amount paid
by the users that only played content provided by them. The second one
states that if a user changes streaming times, then the amount received by
each artist could not change more than the subscription paid by that user.
We explore how the main indices perform with respect to all these axioms,
but our main result in this axiomatic approach to the problem (Theorem 1)
states three characterization results. First, the combination of \textit{%
additivity}, \textit{homogeneity} and \textit{equal individual impact of
similar users} characterizes the \textit{pro-rata} index. Second, the
combination of \textit{additivity}, \textit{homogeneity} and \textit{equal
global impact of users} characterizes the \textit{user-centric} index.
Third, the combination of \textit{additivity} and \textit{homogeneity}
characterizes a family of \textit{weighted} indices that compromise between
the \textit{pro-rata} index and the \textit{user-centric} index, upon
assigning each artist the weighted aggregation of streamings, with the
weight depending on the user and her streaming profile.

We then take two other indirect approaches to deal with streaming problems.
In both of them, streaming problems are associated to other problems that
have already been studied in the literature, and we then import well-known
solutions for those problems to solve streaming problems.

First, we associate each streaming problem with a \textit{cooperative game
with transferable utility}, where the worth of each set of artists is
defined as the amount paid by the users that have only played content from
those artists. Such a game is convex, i.e., the incentives to join a
coalition increase as the coalition grows. It is well known that the core of
a convex game is quite large and we fully characterize it in our Theorem 2.
In words, a core allocation imposes that the amount paid by each user is
divided in any way among the artists listened by this user. Each artist then
receives the sum (over all users) of the corresponding amounts. We show that
the \textit{pro-rata} index does not always guarantee allocations within the
core, whereas the \textit{user-centric} index always does so.

Second, we associate each streaming problem with a \textit{claims problem},
where the users are identified as the \textit{issues} in the claims problem.
Then, problems can be solved in two stages. Either focussing on each issue
(user) first and agents (artists) afterwards, or viceversa. We show (Theorem
3) how the indices we consider for streaming problems can be rationalized as
two-stage claims rules. More precisely, the \textit{pro-rata} index and the 
\textit{user-centric} index can be rationalized as \textit{weighted
proportional rules} (although there is no bijection between this family of
rules and the weighted indices for streaming problems). Furthermore, the
allcation rules these two indices induce can also be described as two-stage
(claims) rules where we first decide the importance of each user and then
the importance of each artist for each user, which is computed as the sum
over all users.

\subsection{Related literature}

The closest paper to ours is Alaei et al. (2022). They also consider the
same streaming platforms that we consider here, which generate revenues by
charging users a subscription fee for unlimited access to the content and
compensate artists through an allocation rule. They also assume that users
are heterogeneous in both their overall consumption and the distribution of
their consumption over different artists, but they model this by referring
to the probability (per usage) that each user type wants to consume the
content of each artist. In our case, we talk about the number of times a
user plays an artist. But, leaving this minor aspect aside, our models are
essentially equivalent. They are also concerned with the pro-rata and
user-centric revenue allocation methods, but focus on characterizing when
these two methods can sustain a set of artists on the platform, as well as
comparing them from both the platform's and the artists' perspectives. In
particular, they show that, despite the cross-subsidization between low- and
high-streaming-volume users, the pro-rata method can be preferred by both
the platform and the artists.\footnote{%
Lei (2023) also obtains that the pro-rata method can outperform the
user-centric method in a simple model with only two artists.} 
More precisely, they show that artists who are predominantly listened to by
users whose overall streaming volume (consumption) is high receive higher
payments with the pro-rata allocation than with the user-centric allocation. 
Consequently, if there is an artist who is extremely popular with users who
have high consumption, 
the pro-rata rule is preferred from the platform's viewpoint (actually, the
platform might not even be profitable with the user-centric rule in this
scenario). More generally, the analysis in Alaei et al. (2022) suggests that 
\textit{\textquotedblleft pro-rata should be the artist's compensation rule
of choice if the users' behavior points to correlation between artists'
popularity and consumption among users and/or if the platform might not be
able to estimate heterogeneity of artists' popularity and consumption among
its users"}.\footnote{%
Alaei et al. (2022) also obtain interesting computational results for this
problem. For instance, they show that the platform's problem of selecting an
optimal portfolio of artists is NP-complete, but develop a polynomial time
approximation scheme for the optimal platform's profit. We do not consider
computational aspects in this paper.} 

Our analysis complements Alaei et al. (2022) in several ways. We provide
normative foundations for both methods, as well as game-theoretical and
other indirect approaches to the problem. And we can safely conclude from
our analysis that there exist powerful arguments to prefer the user-centric
method to the pro-rata method. We acknowledge that the pro-rata method 
made sense when the technology to divide revenue by individual users'
streams did not exist. But that is no longer the case and, consequently, 
a user-centric method that distributes revenue based on each listener's
individual streams seems to be more reasonable (based on the arguments we
provide in our analysis). 
But we are not the first ones suggesting that. Haampland et al. (2022) 
resort to the data provided by a leading music streaming platform in France
(with 427 million streams coming from more than 140,000 unique users during
six months). They obtain from their empirical analysis that, when compared
with the pro-rata method, the user-centric method renders consumers' choices
better aligned with revenue sharing, but also that it favors organic streams
with respect to curated streams, it reduces the superstar phenomenon, and it
moderates the potential bias in curated streams. Dimont (2017) also endorses
switching from a pro-rata method to a user-centric method from a legal
perspective. 
Muikku (2017) finds that as the overall stream count decreases, the revenue
difference between the user-centric and the pro-rata methods increases. The
latter favours artists and tracks, which get the biggest amount of played
streams regardless if they are created by a large number of users with few
plays or a smaller number of users who have played them repeatedly. The
former favours artists with smaller number of streams, especially when the
overall stream count is smaller. However, 
the results depend on the cumulative effects of both individual and user
groups' listening habits. 
Meyn et al. (2023) investigate the monetary consequences of the switch from
the pro-rata method to the user-centric method. 
Using individual-level data from 3,326 participants and data from Spotify
Web API, 
they find a substantial reallocation (mostly shifting revenue from
mainstream to niche genres) of nearly \euro 170 million per year at Spotify,
driven by the song length as well as the payments per listening time. 

Beyond the specific issue of sharing the revenue raised from subscriptions
to (online music) platforms, the literature (at the intersection of several
fields such as economics, management, marketing and law) has addressed other
issues regarding the effect that streaming has on the music industry. For
instance, Aguiar and Waldfogel (2018) find that although streaming displaces
sales, it also displaces music piracy. 
Aguiar and Waldfogel (2021) highlight the market power of some platforms
(especially, Spotify) in the online music market (via their via
platform-operated playlists, which act as promotion channels). 
Datta et al., (2018) analyze the effects of the gradual adoption of
streaming services in the music industry. 
They find that adoption of streaming leads to very large increases in the
quantity and diversity of consumption in the first months after adoption.
And, although the effects attenuate over time, adopters continue playing
substantially more, and more diverse, music. 
Jain and Qian (2021) focus on another major source of revenue for platforms;
namely, advertising, which depends on the number of active customers using
the platforms. 
Their emphasis is on how such sharing incentives are affected by the nature
of competition among various producers, the size of the customer base, and
the type of customers. 
They show that increased producer competition can lead to higher
compensation for the producers, higher content quality, and higher
producers' profits. 
And, more generally, Waldfogel (2017) surveyed the literature dealing with
digitalization and its positive effects in different art industries (not
only music, but also films, literature and television), whereas Rietveld and
Schilling (2020) reviewed the literature (comprising the last three decades)
on platform competition.


Our work also relates to the industrial organization literature dealing with
bundling, which can be traced back to Adams and Yellen (1976). Bundling
products is typically an effective mechamism to increase revenue with
respect to selling products independently. Examples abounded in real life
even before the advent of digital content platforms (e.g., Ginsburgh and
Zang, 2003; Berganti\~{n}os and Moreno-Ternero, 2015). 
Online bundling might nevertheless have a different nature, as it often
occurs in a distribution channel where a downstream firm (retailer) adopts a
bundling strategy involving products made by separate and multiple upstream
firms (manufacturers). This sort of practice not only applies to the
platforms offering entertainment products, such as the ones we analyze here,
but also to other information goods such as online services (e.g., Bhargava,
2013).\footnote{%
Geng et al. (2005) point out how consumers' average value for consuming a
stream of information goods (such as streaming music, or other online
entertainment) declines with the number consumed and provide basic
guidelines for optimal bundling marketing strategies in such a case.} 
The complex relationship between the independent price 
of each product and the bundled price 
renders the problem of sharing the revenue from periodic charges to
unlimited streaming among the participating agents 
a complex one too. Nevertheless, we believe our results could shed light on
that problem.

Our paper (especially, Section 3) is also connected to the sizable axiomatic
literature on resource allocation or, more generally, the axiomatics of
economic design (e.g., Thomson, 2023). A variety of axioms formalizing
principles with ethical or operational appeal for resource allocation have
been introduced in economic theory for more than half a century. Instances
are axioms formalizing no-envy, impartiality, priority, or solidarity (e.g.,
Foley, 1967; Moreno-Ternero and Roemer, 2006; Thomson, 2023). 
They have contributed to our understanding of normative issues concerning
the allocation of goods and services. Our paper extends the scope of this
literature in order to deal with a special and somewhat new type of goods
and services that have arisen in the current digital era. 
But the characterizations we provide in Section 3 are also reminiscent of
other recent characterizations in the axiomatic literature on resource
allocation. 
For instance, Berganti\~{n}os and Moreno-Ternero (2020) analyze the problem
of sharing the (collectively raised) revenues from sportscast.\footnote{%
Dietzenbacher and Kondratev (2023) study the related problem of allocating a
prize endowment among competitors, when their ranking is known.} The input
in their case is the audience matrix indicating the number of viewers of
each game between two clubs participating in a given competition (thus,
similar to the input of the streaming problems we analyze in this paper).
And they also characterize two focal rules to share those (collectively
raised) revenues.\footnote{%
Families of rules compromising among the two rules have also been considered
in the literature (e.g., Berganti\~{n}os and Moreno-Ternero, 2021, 2022).}
Singal et al. (2022) also develop an axiomatic framework to study the
related problem of attribution in online advertising; namely, assessing the
contribution of individual advertiser actions 
to eventual conversion.\footnote{%
Both Berganti\~{n}os and Moreno-Ternero (2020) and Singal et al. (2022) also
explore the game-theoretical approach to solve their problems, as we do in
this paper.} 
Somewhat related, Flores-Szwagrzak and Treibich (2020) introduce an
innovative productivity index 
that disentangles individual from collaborative productivity. This index
defines an individual's productivity score as the sum of her credit over all
the projects she has contributed to. Credit on each project is allocated
proportionally to the score of each teammate and the scores are thus
determined endogenously and simultaneously (solving a fixed point problem).
Martinez and Moreno-Ternero (2022) characterize a family of pandemic
performance indicators arising from a weighted average of the incidence
rate, morbidity rate and mortality rate. Previously, Hougaard and Moulin
(2014) studied how to share the cost of finitely many public goods (items)
among users with different needs. 
They characterized a family of cost ratios based on simple liability
indices, one for each agent and for each item, measuring the relative worth
of this item across agents, and generating cost allocation rules additive in
costs. And, more recently, Gudmundsson et al. (2023) also take the axiomatic
approach to analyze the problem of sharing sequentially triggered losses.
They characterize a class of fixed-fraction rules, which strike a balance
between incentives for accident prevention and fairness to assign
liabilities. Finally, Gon\c{c}alves-Dosantos et al. (2023) also take the
axiomatic approach (but not the game-theoretical, or another indirect
approach) to study general content streaming platforms (such as, for
instance, Netflix). As music streaming platforms have some specific
characteristics that are not necessarily extended to general content
streaming platforms, most of the axioms considered in both papers are of a
different nature. Characterization results are thus genuinely different.%
\footnote{%
As a matter of fact, in all of our characterizations we use axioms that do
not appear in Gon\c{c}alves-Dosantos et al. (2023), whereas in all of the
characterizations they provide, they use some axioms that do not appear in
our paper.} 

To conclude with the introduction, we mention that our paper is also related
to the literature on cooperative game theory. There is a tradition of
analyzing problems involving agents' cooperation with a game-theoretical
approach. Classical instances are the so-called airport problems (e.g.,
Littlechild and Owen, 1973), in which the cost of a runway has to be shared
among different types of airplanes, bankruptcy problems from the Talmud
(e.g., Aumann and Maschler, 1985), telecommunications problems and the
rerouting of international telephone calls (e.g., van den Nouweland et al.,
1996), or modern transportation systems (e.g., Krajewska et al., 2008). One
of the approaches we take in this paper (Section 4) is precisely this one. 
The (cooperative) game we associate to streaming problems is convex and,
thus, it has a large core encompassing allocations guaranteeing
participation constraints. We fully characterize such a core and show that
the pro-rata method can generate allocations that violate these constraints.
We also show that the resulting axiom of core-selection is key to
characterize the user-centric method. The combination of these two results
is another powerful argument we provide to support the user-centric method
with respect to the pro-rata method.

The game-theoretical approach is an indirect way of solving streaming
problems. Another indirect approach suggests to solve them via associating a
claims problem (rather than a cooperative game). The problem of adjudicating
conflicting claims (in short, claims problem) models a basic situation in
which an endowment is allocated among agents who have claims on it, and the
available amount is not enough to fully honor all claims. This is a classic
problem that can be traced back to ancient sources, such as Aristotle and
the Talmud, although is formal treatment is somewhat recent (e.g., O'Neill,
1982; Thomson, 2019). Ju et al. (2007) generalize these problems to account
for multiple issues.\footnote{%
Cs\'{o}ka and Herings (2018, 2021) also studied recently a different
generalization of claims problems to deal with financial networks, as
pioneered by Eisenberg and Noe (2001). See also Calleja and Llerena (2023).}
We show that some of the two-stage rules from generalized claims problems
can rationalize some of the allocation rules we consider to solve streaming
problems.

The rest of the paper is organized as follows. In Section 2, we introduce
our model and main concepts for our analysis of streaming problems. 
In Section 3, we present the axiomatic approach to our problems including
the main characterization results we obtain. In Section 4, we explore the
game-theoretical approach to our problem. In Section 5, we explore another
indirect approach to solve our problems based on claims problems. Section 6
concludes. Most of the proofs have been postponed to an appendix.

\section{Streaming problems}

Let $N=\left\{ 1,...,n\right\} $ denote a set of artists and $M=\left\{
1,...,m\right\} $ a set of users. For each pair $i\in N,j\in M$, let $t_{ij}$
denote the times user $j$ played (via streaming) contents uploaded by artist 
$i$ in the platform, briefly streams, during a certain period of time (e.g.,
month). 
Let $t=\left( t_{ij}\right) _{i\in N,j\in M}$ denote the corresponding
matrix encompassing all playing times. A \textbf{streaming problem} is $%
P=\left( N,M,t\right) $.\footnote{%
As mentioned at the Introduction, and for ease of notation, 
we normalize the amount paid by each user to 1. Thus, the amount to be
divided among artists in a problem $\left( N,M,t\right) $ is just $m$, the
number of users.} The set of problems so defined is denoted by $\mathcal{P}$%
. 
For each $j\in M,$ we denote by $t^{-j}$ the matrix obtained from $t$ by
removing the column corresponding to user $j$.

For each artist $i\in N$, 
\begin{equation*}
T_{i}\left( N,M,t\right) =\sum_{j\in M}t_{ij},
\end{equation*}%
denotes the total times $i$ was played. Likewise, for each user $j\in M$ 
\begin{equation*}
T^{j}\left( N,M,t\right) =\sum_{i\in N}t_{ij},
\end{equation*}%
denotes the total times $j$ played content. 

We define the set of fans of each artist as the set of users who have played
content from the artist at least once. Formally, $F:N\rightarrow M$ is such
that for each $i\in N,$ 
\begin{equation*}
F_{i}\left( N,M,t\right) =\left\{ j\in M:t_{ij}>0\right\} .
\end{equation*}

Similarly, we define the list of artists of a user as those from which the
user has played content at least once. Formally, $L:M\rightarrow N$ such
that for each $j\in M,$ 
\begin{equation*}
L^{j}\left( N,M,t\right) =\left\{ i\in N:t_{ij}>0\right\} .
\end{equation*}

The profile of user $j$ is defined as the streaming vector associated to
such a user. Namely, 
\begin{equation*}
t_{.j}\left( N,M,t\right) =\left( t_{ij}\right) _{i\in N}.
\end{equation*}


When no confusion arises we write $T_{i}$ instead of $T_{i}\left(
N,M,t\right) $, $T^{j}$ instead of $T^{j}\left( N,M,t\right) ,$ $F_{i}$
instead of $F_{i}\left( N,M,t\right) ,$ $L^{j}$ instead of $L^{j}\left(
N,M,t\right) ,$ and $t_{.j}$ instead of $t_{.j}\left( N,M,t\right) .$

We illustrate our model with a basic example that will surface throughout
our ensuing analysis. Assume two users ($a,b$) join a platform to listen to
their favorite artists (1,2). Assume artist $1$ is listened by user $a$,
whereas artist $2$ is listened by user $b$. 
This situation can be included in our theoretical model as follows.

\begin{example}
\label{ex 2,2,a} Let $N=\left\{ 1,2\right\} ,$ $M=\left\{ a,b\right\} ,$ and 
\begin{equation*}
t=\left( 
\begin{array}{cc}
10 & 0 \\ 
0 & 90%
\end{array}%
\right) .
\end{equation*}
\end{example}

\medskip 

A popularity \textbf{index} $\left( I\right) $ for streaming problems is a
mapping that measures the importance of each artist in each problem.
Formally, $I:\mathcal{P}\to \mathbb{R}_{+}^{N}$ and, for each pair $i,j\in N$%
, $I_{i}\left( N,M,t\right)\ge I_{j}\left( N,M,t\right)$ if and only if $i$
is at least as important as $j$ at problem $\left(N,M,t\right)$. 
We assume that $\sum\limits_{i\in N}I_{i}\left( N,M,t\right) >0$. 

The reward received by each artist $i\in N$ from the revenues generated in
each problem ($m$ because the amount paid by each user has been normalized
to 1) is based on the importance of that artist in that problem. Formally, 
\begin{equation*}
R_{i}^{I}\left( N,M,t\right) =\frac{I_{i}\left( N,M,t\right) }{%
\sum\limits_{i^{\prime }\in N}I_{i^{\prime }}\left( N,M,t\right) }m.
\end{equation*}


Note that any positive linear transformation of a given index generates the
same allocation of rewards. Formally, for each $\lambda>0$ and each index $I$%
, $R^{\lambda I}\equiv R^{I}$. Thus, in what follows, we shall slightly
abuse language to identify an index with all its positive linear
transformations too.

The index used in most of the platforms is the so called \textbf{pro-rata}
index, which simply measures importance by the total number of streams.
Formally, for each problem $
\left( N,M,t\right) \in \mathcal{P}$ and each artist $i\in N,$ 
\begin{equation*}
P_{i}\left( N,M,t\right) =T_{i}=\sum_{j\in M}t_{ij}.
\end{equation*}

Thus, the amount received by each artist $i\in N$ under $P$ is

\begin{equation*}
R^P_{i}\left( N,M,t\right)=\frac{T_{i}}{\sum\limits_{j\in N}T_{j}}m.
\end{equation*}


Another basic index is the so called \textbf{user-centric} index. All users
have the same importance (normalized to 1). The importance of each user is
divided among the artists listened by this user proportionally to the total
number of streams. Then, the importance of each artist is the sum, over all
users, of the importance of the artist given by each user. 
Formally, for each problem $\left( N,M,t\right) $ and each artist $i\in N,$ 
\begin{equation*}
U_{i}\left( N,M,t\right) =\sum_{j\in M}\frac{t_{ij}}{T^{j}}.
\end{equation*}%
%
%
%
%
%
%
%
%
%
%
%
%
%
%
%
As 
\begin{equation*}
\sum_{i\in N}\sum_{j\in M}\frac{t_{ij}}{T^{j}}=\sum_{j\in M}\sum_{i\in N}%
\frac{t_{ij}}{T^{j}}=\sum_{j\in M}1=m,
\end{equation*}%
it follows that the amount received by each artist $i\in N$ from this index
is precisely 
\begin{equation*}
R_{i}^{U}\left( N,M,t\right) =U_{i}\left( N,M,t\right) .
\end{equation*}

In Example \ref{ex 2,2,a}, we then have the following: 
\begin{equation*}
\begin{tabular}{cccc}
& $R_{i}^{P}\left( N,M,t\right) $ & $R_{i}^{U}\left( N,M,t\right) $ & $T_{i}$
\\ 
1 & 0.2 & 1 & 10 \\ 
2 & 1.8 & 1 & 90%
\end{tabular}%
\end{equation*}

That is, pro-rata gives to artist 2 much more than to artist 1, whereas the
user-centric gives the same to both artists (which sounds more reasonable in
this example). This might illustrate why some platforms are moving from
pro-rata to user-centric.\footnote{%
For instance, SoundCloud did so in 2021 (e.g., Ingham, 2021) whereas Deezer
started an online campaign promoting user-centric (e.g., Deezer, 2019).
Muikku (2017), Alaei et al. (2022), Haampland et al. (2022), and Meyn et al.
(2023), among others, have compared the two indices on different grounds.} 

There is also room for alternative payment schemes beyond the previous two.
For instance, Meyn et al.., (2022) write \textquotedblleft remuneration
based on quality ratings, or a combination of user-centric and pro-rata
remuneration\textquotedblright . And, also, \textquotedblleft in addition to 
pro-rata and user-centric, the distribution parameters (e.g., the unit of
distribution) must be investigated further\textquotedblright . Partly
accounting for this motivation, we now introduce a family of indices that
allow us to consider alternative payment schemes compromising between the
previous two.

A weight system is a function $\omega :M\times \mathbb{N}_{+}^{N}\rightarrow 
\mathbb{R}$ such that for each $j\in M$ and each $x\in \mathbb{N}_{+}^{N},$ $%
\omega \left( j,x\right) >0.$ For each weight system $\omega $, its \textbf{%
weighted index} $I^{\omega}$ is defined so that for each $\left(
N,M,t\right) $ and each $i\in N,$ 
\begin{equation*}
I_{i}^{\omega}\left( N,M,t\right) =\sum_{j\in M}w\left( j,t._{j}\right)
t_{ij}.
\end{equation*}

The index of each artist is obtained as the sum, over all users, of the
streams of the user weighted by a factor that depends on the user (allowing,
for instance, for popular users to have more importance than unknown users)
and the streaming profile of the user (allowing, for instance, for more
active users to have more importance than less active users). The pro-rata
index is the weighted index associated with $\omega\left( j,x\right) =1$ for
each $j\in M$, and each $x\in \mathbb{N}_{+}^{N}\ $(namely, the users and
the profile do not matter) and the user-centric index is the weighted index
associated with $\omega \left( j,x\right) =\frac{1}{\sum\limits_{i\in N}x_{i}%
}$ for each $j\in M$, and each $x\in \mathbb{R}_{+}^{N}$ (namely, the user
does not matter but the profile matters).

Notice that under pro-rata the importance of each user to all artists,
namely $\sum\limits_{i\in N}w\left( j,t._{j}\right) t_{ij}=T^{j},$ is
proportional to his/her total streams. Under user-centric all users have the
same importance, namely $\sum\limits_{i\in N}w\left( j,t._{j}\right)
t_{ij}=1 $. We can consider weight systems reflecting that all users have a
minimum importance and also that users with more streamings contribute more,
but with an upper bound. For instance, for each $j\in M$,

\begin{equation*}
w\left( j,x\right) =\left\{ 
\begin{tabular}{ll}
$\frac{1}{\sum\limits_{i=1}^{n}x_{i}}$ & if $\sum\limits_{i=1}^{n}x_{i}\leq
\alpha $, \\ 
$\frac{1}{\alpha }$ & if $\alpha <\sum\limits_{i=1}^{n}x_{i}\leq \beta $, \\ 
$\frac{\beta }{\alpha \sum\limits_{i=1}^{n}x_{i}}$ & if $\sum%
\limits_{i=1}^{n}x_{i}>\beta $.%
\end{tabular}%
\right.
\end{equation*}

Notice that $\omega\left( j,x\right) $ does not depend on $j.$ Thus, we only
distinguish users through the streaming times. 
More precisely, all users with a number of streams below $\alpha $ have the
same importance as in the user-centric index. Namely, for each $j\in M$, 
\begin{equation*}
\sum_{i\in N}w\left( j,t_{.j}\right) t_{ij}=1.
\end{equation*}

Users with a number of streams between $\alpha $ and $\beta $ have a
proportional importance to the streams.Namely, for each $j\in M$,%
\begin{equation*}
\sum_{i\in N}w\left( j,t_{.j}\right) t_{ij}=\frac{T^{j}}{\alpha }.
\end{equation*}

Finally, users with a number of streams above $\beta $ have the same
importance, given by the thresholds $\alpha $ and $\beta $. Namely, for each 
$j\in M$, 
\begin{equation*}
\sum_{i\in N}w\left( j,t_{.j}\right) t_{ij}=\frac{\beta }{\alpha }.
\end{equation*}

We now revisit Example \ref{ex 2,2,a}, adding a new user ($c$) with $5$
streams for artist $1$ and $35$ streams for artist $2$. Besides, let $\alpha
=20$ and $\beta =60.$ Then, the weigthed index $I^{\alpha ,\beta }$ defined
as above is 
\begin{eqnarray*}
I_{1}^{\alpha ,\beta }\left( N,M,t\right) &=&\left( \frac{1}{10}\right)
10+\left( \frac{60}{20\ast 90}\right) 0+\left( \frac{1}{20}\right) 5=1.25%
\text{ and } \\
I_{2}^{\alpha ,\beta }\left( N,M,t\right) &=&\left( \frac{1}{10}\right)
0+\left( \frac{60}{20\ast 90}\right) 90+\left( \frac{1}{20}\right) 35=4.75.
\end{eqnarray*}

In this case, $P\left( N,M,t\right) =\left( 15,125\right) $ whereas $U\left(
N,M,t\right) =\left( 1.125,1.875\right)$. Then, the rewards induced for
artists are 
\begin{equation*}
\begin{tabular}{cccc}
& $R_{i}^{P}\left( N,M,t\right) $ & $R_{i}^{U}\left( N,M,t\right) $ & $%
R_{i}^{I^{\alpha ,\beta }}\left( N,M,t\right) $ \\ 
1 & 0.3 & 1.1 & 0.6 \\ 
2 & 2.7 & 1.9 & 2.4%
\end{tabular}%
\end{equation*}

We observe that $I^{\alpha ,\beta }$ yields an allocation in between those
the pro-rata and user-centric yield.

\section{An axiomatic approach}

In this section, we take the axiomatic approach to solve streaming problems.
That is, we formalize axioms of indices that model principles with normative
(ethical or operational) appeal. Some of them will echo the concern that
artists are paid fairly.\footnote{%
Haamland et al., (2022) provide an interesting discussion about this issue.}
Some others are inspired from related discussions in the music industry or
from the literature on resource allocation. \bigskip

To introduce the first axiom, which is standard in axiomatic work, assume
that each user has played artist $i$ a certain times more than artist $%
i^{\prime }.$ Then, the index should preserve that ratio. Formally,

\textbf{Homogeneity.} 
For each $\left( N,M,t\right) \in \mathcal{P} $, each pair $i,i^{\prime }\in
N,$ and each $\lambda \geq 0$ such that $t_{ij}=\lambda t_{i^{\prime }j}$
for all $j\in M,$ $\ $ 
\begin{equation*}
I_{i}\left( N,M,t\right) =\lambda I_{i^{\prime }}\left( N,M,t\right) .
\end{equation*}

The second axiom is also a standard axiom in resource allocation and says
that if we can divide a problem in the sum of two smaller problems, then the
solution to the original problem should be the sum of the solutions in the
two smaller problems. The intuition in streaming problems is the following.
Suppose that a platform operates in several countries. Then, we can reward
artists in two ways. First, we consider all countries in the same market and
we allocate artists according with the streams in all countries. Second, we
consider each country as a different market. Thus, each artist receives an
allocation in each country according with the streams in this country. The
total allocation to an artist is the sum over all countries. Additivity says
that both ways should coincide. Formally,

\textbf{Additivity. }
For each trio $\left( N,M^{1},t^{1}\right), \left( N,M^{2},t^{2}\right) ,
\left( N,M,t\right) \in \mathcal{P}$ such that $M=M^{1}\cup M^{2}$, $%
t_{ij}=t_{ij}^{1}$ when $j\in M^{1}$ and $t_{ij}=t_{ij}^{2}$ when $j\in
M^{2},$ 
\begin{equation*}
I\left( N,M,t\right) =I\left( N,M^{1},t^{1}\right) +I\left(
N,M^{2},t^{2}\right) .
\end{equation*}

We now consider two alternative forms of modeling the impact of extra users. 


Assume that two users (say, $j$ and $j^{\prime })$ have the same streams on
artist $i.$ We can then consider that both users are similar for artist $i$.
Then, both users should have the same impact over this artist.\footnote{%
A rationale for this axiom is provided by Page and Safir (2018a), who write
\textquotedblleft some listeners may find it inequitable that a given track
be awarded significantly different per-stream values\textquotedblright . 
} Formally,

\textbf{Equal individual impact of similar users}. 
For each $\left( N,M,t\right) \in \mathcal{P} $, each $i\in N,$ and each
pair $j,j^{\prime }\in M$ such that $t_{ij}=t_{ij^{\prime }}$, 
\begin{equation*}
I_{i}\left( N,M\backslash \left\{ j\right\} ,t^{-j}\right) =I_{i}\left(
N,M\backslash \left\{ j^{\prime }\right\} ,t^{-j^{\prime }}\right) .
\end{equation*}

\bigskip

It is also often argued that as all users pay the same, all users should
have the same impact on the index. We formalize this idea as follows:

\textbf{Equal global impact of users. } 
For each $\left( N,M,t\right) \in \mathcal{P} $ and each pair $j,j^{\prime
}\in M,$ 
\begin{equation*}
\sum_{i\in N}I_{i}\left( N,M\backslash \left\{ j\right\} ,t^{-j}\right)
=\sum_{i\in N}I_{i}\left( N,M\backslash \left\{ j^{\prime }\right\}
,t^{-j^{\prime }}\right) .
\end{equation*}

The next result states the characterizations we obtain combining the
previous axioms.%



\begin{theorem}
\label{char family} The following statements hold:

$\left( a\right) $ An index satisfies \textit{homogeneity}, and \textit{%
additivity} if and only if it is a weighted index.

$\left( b\right) $ An index satisfies \textit{homogeneity}, \textit{%
additivity}, and \textit{equal individual impact of similar users} if and
only if it is the pro-rata index.

$\left( c\right) $ An index satisfies \textit{homogeneity}, \textit{%
additivity}, and \textit{equal global impact of users} if and only if it is
the user-centric index.
\end{theorem}


Theorem \ref{char family}, whose proof is postponed to the appendix,
provides normative grounds for the indices considered above. Statement $%
\left( a\right) $ provides the characterization of the family of indices
satisfying the two basic axioms. Pro-rata and user-centric arise from the
family by imposing an additional axiom. \textit{Equal individual impact of
similar users} for the former and \textit{equal global impact of users} for
the latter. Both axioms reflect different ways to address users and streams. 
The former focuses on streams while stating, roughly speaking, that all
streams are equally important, independently of who has produced the
streams. The latter focuses on users while stating, roughly speaking, that
all users are equally important, independently of the streams of each user.
Thus, we can see pro-rata and user-centric as two somewhat polar weighted
indices. The first one \textquotedblleft equalizes\textquotedblright\
streams, whereas the second one \textquotedblleft
equalizes\textquotedblright\ users.

\bigskip

We conclude this section complementing the axiomatic analysis from Theorem %
\ref{char family} upon formalizing two axioms that capture ideas that have
recently received attention in the literature on the music industry. We
shall see that, although Theorem \ref{char family} puts both the
user-centric index and the pro-rata index on a similar foot, the following
analysis (with respect to the new two axioms) favors the user-centric index.

Haampland et al. (2022) write the following: \textquotedblleft such a
payment system can be perceived to be quite unfair since individual users'
payments are not aligned with their actual musical
preferences.\textquotedblright\ Similarly, Meyn et al. (2023) argue that
users could \textquotedblleft prefer that their payment only goes towards
the content they use\textquotedblright . Thus, we consider an axiom that
says that given a set of users $C$ and the set of artists $A$ listened by
those users, the amount received by the artists in $A$ should be, at least,
the amount paid by users in $C.$ Formally,

\textbf{Reasonable lower bound. }
For each $\left( N,M,t\right) \in \mathcal{P}$ and each $C\subset M$, let $%
L^{C}=\bigcup\limits_{j\in C}L^{j}.$ Then, 
\begin{equation*}
\sum_{i\in L^{C}}\frac{I_{i}\left( N,M,t\right) }{\sum\limits_{i^{\prime
}\in N}I_{i^{\prime }}\left( N,M,t\right) }m\geq \left\vert C\right\vert .
\end{equation*}


On another matter, Meyn et al. (2023) write the following: \textquotedblleft
some artists have asked their listeners to play their music silently while
asleep to generate a larger share of remuneration\textquotedblright . Dimont
(2017) calls this phenomenon \textit{click fraud} and discuss it giving some
real examples. We first notice that it is not possible to know if some user
is making click fraud or not (namely, if he/she is listening the song or not
when playing it). Thus, it seems reasonable that if some song is played, the
artists should be rewarded for it. Nevertheless, it should not be rewarded
too much. For instance, it does not seem reasonable that if a user pays $%
\$10 $ and decides to listen to an artist many times, then the amount
received by this artist increases more than $\$50$. Otherwise, artists would
have incentives to create fictitious users listening artist's songs all the
day. We formalize this axiom by saying that if a user changes his/her
streaming times, then the amount received by each artist could not change
more that the subscription paid by the user. Formally,

\textbf{Click-fraud-proofness. }
Let $\left( N,M,t\right) ,\left( N,M,t^{\prime }\right) \in \mathcal{P}$ and 
$j\in M$ be such that $t_{ij^{\prime }}=t_{ij^{\prime }}^{\prime }$ for all $%
j^{\prime }\in M\backslash \left\{ j\right\} $ and $i\in N$. Then, for all $%
i\in N,$ 
\begin{equation*}
\left\vert \frac{I_{i}\left( N,M,t\right) }{\sum\limits_{i^{\prime }\in
N}I_{i^{\prime }}\left( N,M,t\right) }m-\frac{I_{i}\left( N,M,t^{\prime
}\right) }{\sum\limits_{i^{\prime }\in N}I_{i^{\prime }}\left( N,M,t^{\prime
}\right) }m\right\vert \leq 1.
\end{equation*}

\bigskip

The user-centric index satisfies the previous two axioms, whereas pro-rata
satisfies none of them (see the appendix for the details).


\section{A game-theoretical approach}

A standard approach to solve resource allocation problems is through
cooperative game theory. That is, given a resource allocation problem we
associate a cooperative game, we compute a cooperative solution (for
instance the core), and finally we compute the allocation induced by the
cooperative solution in the original problem. We shall follow that approach
in this section to solve our streaming problems.

A \textbf{cooperative game with transferable utility}, briefly a \textbf{TU
game}, is a pair $\left( N,v\right) $, where $N$ denotes a set of agents and 
$v:2^{N}\rightarrow \mathbb{R}$ satisfies $v\left( \varnothing \right) =0.$
The \textbf{core} of a cooperative game is defined as the set of feasible
payoff vectors, for which no coalition can improve upon. Formally, for each
game $\left( N,v\right) $, 
\begin{equation*}
C\left( N,v\right) =\left\{ x=(x_{i})_{i\in N}\text{: }\sum_{i\in
N}x_{i}=v\left( N\right) \text{ and }\sum_{i\in S}x_{i}\geq v(S)\text{ for
each }S\subset N\right\} .
\end{equation*}

\bigskip

We now associate with each streaming problem $\left( N,M,t\right) $ a $TU$
game $\left( N,v_{\left( N,M,t\right)} \right) $ where the set of agents of
the cooperative game is the set of artists. Given $S\subset N$ we define $%
v_{\left( N,M,t\right)} \left( S\right) $ as the amount paid by the users
that have only listened to artists in $S$.\footnote{%
The definition of $\left( N,v_{\left( N,M,t\right)} \right) $ is similar to
the one used in the so-called museum pass problem (e.g., Ginsburg and Zang,
2003; Berganti\~{n}os and Moreno-Ternero, 2015).} 
Formally, 
\begin{equation*}
v_{\left( N,M,t\right)} \left( S\right) =\left\vert \left\{ j\in
M:L^{j}\subset S\right\} \right\vert .
\end{equation*}

When no confusion arises we write $v$ instead of $v_{\left( N,M,t\right)} $.
Note that this game is based on a pessimistic premise. To wit, it is
reasonable to assume that if artists do not consider themselves to be well
paid, they might leave the platform.\footnote{%
Kanye West, a highly popular artist, recently launched his own streaming
service (e.g., Meyn et al., 2023).} The issue is to estimate the revenues
seceding artists might obtain after leaving the platform (and creating a new
one). 
Our estimation is that just all users that listened only to the seceding
artists will leave the platform to join the new one promoted by them. We
acknowledge that this could be 
a quite pessimistic stance. For instance, consider a user whose $99\%$ of
streams are devoted to artists within the seceding group ($S$) and only $1\%$
to other artists. It seems plausible to conjecture that this user would
leave the platform, whereas our assumption on $v(S)$ does not consider it.


It is not difficult to show that $v$ is a convex game, i.e., the incentives
to join a coalition increase as the coalition grows (or, more formally, its
characteristic function is supermodular). It is well known that the core of
a convex game is quite large. The next result actually characterizes all the
allocations within the core of this game. In words, they must satisfy the
following: the amount paid by each user (which we have normalized to 1) is
divided in any way among the artists listened by this user. Each artist then
receives the sum (over all users) of the corresponding amounts. Formally,
for each $\left( N,M,t\right) $ we define the following set of allocations:

\begin{equation*}
A\left( N,M,t\right) =\left\{ 
\begin{array}{c}
x\in \mathbb{R}^{N}:x=\sum\limits_{j\in M}x^{j}\text{ where for each }j\in M
\\ 
x^{j}\in \mathbb{R}^{N}, \\ 
x_{i}^{j}=0\text{ for each }i\in N\backslash L^{j}\text{,} \\ 
x_{i}^{j}\geq 0\text{ for each }i\in L^{j}\text{ and} \\ 
\sum\limits_{i\in N}x_{i}^{j}=1.%
\end{array}%
\right\} .
\end{equation*}

\bigskip

\begin{theorem}
\label{core all}For each $\left( N,M,t\right)\in \mathcal{P}$, 
\begin{equation*}
C\left( N,v\right) =A\left( N,M,t\right) .
\end{equation*}
\end{theorem}


The proof of Theorem \ref{core all} can be found in the appendix.


We note that the rewards induced by the pro-rata index could be outside the
core. To do so, consider for instance Example \ref{ex 2,2,a} and $S=\left\{
1\right\} .$ Then, 
$R_{1}^{P}\left( N,M,t\right) =\frac{20}{100}<1=v\left( S\right)$. 

On the other hand, by definition, the user-centric index yields for each
problem $\left( N,M,t\right)$ an allocation that belongs to $A\left(
N,M,t\right)$. Thus, by Theorem \ref{core all}, the rewards induced by the
user-centric index always belong to the core. 
More generally, we introduce the axiom stating that the rewards generated by
an index should always lie within the core of the associated cooperative
game. Formally,

\textbf{Core selection. }For each $\left( N,M,t\right)\in \mathcal{P} $, $%
R^{I}\left( N,M,t\right) \in C\left( N,v\right) $.


As the next result states, \textit{core selection} actually characterizes
the user-centric index, when combined with the axioms of \textit{homogeneity}
and \textit{additivity} (already introduced in the previous section),
provided we restrict to the following relevant subdomain of streaming
problems.

More precisely, let $\mathcal{P}^{\ast }$ be the set of all problems (with
at least three users) where no user has played content from all the artists.
Namely, 
\begin{equation*}
\mathcal{P}^{\ast }=\left\{ \left( N,M,t\right) \in \mathcal{P}:\left\vert
M\right\vert \geq 3\text{ and }L^{j}\left( N,M,t\right) \neq N\text{ for all 
}j\in M\right\} .
\end{equation*}

\begin{theorem}
\label{char user-centric}In the domain $\mathcal{P}^{\ast }$, an index
satisfies \textit{homogeneity}, \textit{additivity}, and \textit{core
selection} if and only if it is the user-centric index.
\end{theorem}


The proof of Theorem \ref{char user-centric} can be found in the appendix.
Note that $\mathcal{P}^{\ast }$ is not a very restrictive domain for
real-life platforms.\footnote{%
Obviously, all platforms have at least three users. And a user can play at
most $89280$ streams per month (assuming 30 seconds to qualify as a valid
streaming, and the user is somehow playing 24/7 those shortest valid
streamings), which is typically below the number of artists platforms have
in their catalogue.} Nevertheless, one could extend Theorem \ref{char
user-centric} to the full domain $\mathcal{P}$ upon simply adding an axiom
of \textit{independence of null artists}, stating that removing an artist
with no streamings does not change the value of the index for the remaining
artists.


\bigskip

Based on the analysis in this section, we can safely state that the
game-theoretical approach favors the user-centric index with respect to the
pro-rata index. 

\section{A claims approach}

In this section, we consider another (indirect) approach to solve streaming
problems, based on claims problems. Claims problems refer to an amount of a
homogeneous and infinitely divisible good (e.g., money) to be divided among
a set of agents, who have claims on the good. This is certainly the case of
the canonical and well-known bankruptcy problem.\footnote{%
Although this problem can be traced back to Aristotle and the Talmud, the
seminal contribution is O'Neill (1982). Thomson (2019) is an excellent
survey with an extensive treatment of the sizable literature emanating from
that seminal contribution.} 
Formally, a \textbf{bankruptcy problem} is a triple $\left( N,c,E\right) $
where $N$ is the set of agents, $c\in \mathbb{R}_{+}^{N}$ is the claims
vector and $E\in \mathbb{R}_{+}$ is the amount to be divided. It is assumed
that $\sum\limits_{i\in N}c_{i}\geq E$. A rule is a function $R$ assigning
to each bankruptcy problem $\left( N,c,E\right) $ a vector $R\left(
N,c,E\right) \in \mathbb{R}^{N}$ such that for each $i\in N$, $0\leq
R_{i}\left( N,c,E\right) \leq c_{i}$ and $\sum\limits_{i\in N}R_{i}\left(
N,c,E\right) =E$. Some popular rules are the \textbf{proportional rule},
which yields awards proportionally to claims, and the \textbf{constrained
equal awards rule}, which equalizes the amount received by each agent as
much as possible. Formally, for each $(N,c,E)$ and each $i\in N,$ 
\begin{equation*}
P_{i}(N,c,E)=\frac{c_{i}}{\sum\limits_{i^{\prime }\in N}c_{i^{\prime }}}\,E, 
\text{ and }
\end{equation*}
\begin{equation*}
CEA_{i}(N,c,E)=\min \{\lambda ,c_{i}\}
\end{equation*}%
where $\lambda $ satisfies $\sum\limits_{i\in N}\min \{\lambda ,c_{i}\}=E$.

We now associate with each streaming problem $\left( N,M,t\right) $ a
bankruptcy problem $\left( N,c\left( N,M,t\right) ,E_{\left( N,M,t\right)
}\right) $ where $c_{i}\left( N,M,t\right) =T_{i}$, and $E\left(
N,M,t\right) =m.$ Notice that $R_{i}^{P}\left( N,M,t\right) ,$ the amount
received by artist $i$ under the pro-rata index, coincides with $P_{i}\left(
N,c,m\right) ,$ the amount received by agent $i$ in the associated
bankruptcy problem under the proportional rule. Besides, $U_{i}\left(
N,M,t\right) ,$ the amount received by artist $i$ under the user-centric
index, can not be computed through $\left( N,T,m\right) $ because it depends
on numbers ($t_{ij})$ that do not appear in $\left( N,T,m\right) .$ We now
present an extension of bankruptcy problems, following Ju et al., (2007),
that allows us to use the values $t_{ij}$.


A\textbf{\ claims problem} is a tuple $\left( N,K,c,E\right) $ where $N$ is
a set of agents, $K$ is a set of issues, $c=(c_{ij})_{i\in N,j\in K}\ $where
for all $i\in N$ and $j\in K,$ $c_{ij}\geq 0$ denotes the characteristic of
agent $i$ on issue $j$, and $E\in \mathbb{R}_{+}$ is the amount of a
homogeneous and infinitely divisible good to be divided. A rule is a
function $R$ assigning to each claims problem $\left( N,K,c,E\right) $ a
vector $R\left( N,K,c,E\right) \in \mathbb{R}^{N}.$


We now associate with each streaming problem $\left( N,M,t\right) $ a claims
problem $\left( N,K\left( N,M,t\right) ,c\left( N,M,t\right) ,E\left(
N,M,t\right) \right) $ where $K\left( N,M,t\right) =M,$ $c\left(
N,M,t\right) =t$, and $E\left( N,M,t\right) =m.$


Ju et al., (2007) characterize several families of rules.\footnote{%
The do so mostly thanks to the axiom of reallocation-proofness, which does
not have a parallel in this paper.} One of the families is formally defined
as follows.

For each claims problem $\left( N,K,c,E\right) $ and each $j\in K$, let $%
c_{.j}=(c_{ij})_{i\in N}$ and $C^{j}=\sum\limits_{i\in N}c_{ij}.$ A weight
function is a function $\omega :\mathbb{R}_{+}^{K}\times \mathbb{R}%
_{+}\rightarrow \mathbb{R}_{+}^{K}$, which assigns a probability
distribution $\omega (x,y)$ (namely, $0\leq w_{j}\left( x,y\right) \leq 1$
for all $j\in K$ and $\sum\limits_{j\in K}w_{j}\left( x,y\right) =1)$. The 
\textbf{weighted proportional rule} associated to $\omega $ assigns, for
each problem $\left( N,K,c,E\right) $ and each $i\in N$ the amount 
\begin{equation*}
P_{i}^{\omega }\left( N,K,c,E\right) =\sum_{j\in K}\frac{c_{ij}}{C^{j}}%
\omega _{j}(\left( C^{j}\right) _{j\in K},E)E.
\end{equation*}

In words, rule $P^{\omega }$ first applies the proportional rule to each
single-dimensional sub-problem $\left( N,\left\{ j\right\} ,c_{.j},E\right) $
and then it takes the weighted average of the solutions to the sub-problems
according to the vector of weights $\omega _{j}(\left( C^{j}\right) _{j\in
K},E)$.

Calleja et al., (2005) introduce multi-issue allocation situations, which
are a particular case of claims problems.\footnote{%
Actually the tuple $\left( N,K,c,E\right) $ is defined as in claims problems
but some additional constraints on $c$, $E$ and the definition of a rule are
added. As our results are not affected by that, we avoit the details.}
Berganti\~{n}os et al., (2010, 2011, 2018) consider two-stage rules for
claims problems where in the first stage the endowment is divided among the
issues and in the second stage the amount assigned to each issue is divided
among the agents. The final amount received by each agent is the sum over
all issues.

Formally, let $\psi $ and $\phi $ be two bankruptcy rules. The\textbf{\
two-stage rule} $R^{\psi ,\phi }$ is the claims rule obtained from the
following two-stage procedure:

\begin{enumerate}
\item First stage. We consider the bankruptcy problem among the issues $%
(K,c^{K},E)$, where $c^{K}=(c_{j}^{K})_{j\in K}$ and for each $j\in K$, $%
c_{j}^{K}=\sum\limits_{i\in N}c_{ji}$ . We compute $\psi (K,c^{K},E).$

\item Second stage. For each $j\in K$, we consider the bankruptcy problem $%
(N,\psi _{j}\left( K,E,c^{K}\right) ,c_{.j})$. We compute $\phi
(N,c_{.j},\psi _{j}\left( K,c^{K},E\right) )$.
\end{enumerate}

Thus, for each $i\in N$, 
\begin{equation*}
R_{i}^{\psi ,\phi }(N,K,C,E)=\sum_{j\in K}\phi _{i}\left( N,c_{.j},\psi
_{j}\left( K,c^{K},E\right) \right) .
\end{equation*}

Moreno-Ternero (2009) and Berganti\~{n}os et al., (2010) study the two-stage
rule where the proportional rule is used in both stages (namely, $\psi =\phi
=P).$ Berganti\~{n}os et al. (2011) study the two-stage rule where the
constrained equal awards rule is used in both stages (namely, $\psi =\phi
=CEA).$ Berganti\~{n}os et al. (2018) study the two-stage rule where the
constrained equal awards rule is used in the first stage and the
proportional rule in the second stage (namely, $\psi =CEA$ and $\phi =P).$
But all those papers take the axiomatic approach inspired on the literature
of bankruptcy problems, whch is unrelated to the axiomatic study of this
paper.

We obtain the following relationships between the pro-rata index and the
user-centric index and some of the rules from the literature on claims
problems.

\begin{theorem}
\label{claims results} Let $\left( N,M,t\right) $ be a streaming problem and 
$\left( N,K,c,E\right) $ be the associated claims problem. Then,

$\left( a\right) $ There exist weight functions $\omega ^{P}$ and $\omega
^{U}$ such that $R^{P}\left( N,M,t\right) =P^{\omega ^{P}}\left(
N,K,c,E\right) $ and $R^{U}\left( N,M,t\right) =P^{\omega ^{U}}(N,K,c,E).$ 

$\left( b\right) $ $R^{P}\left( N,M,t\right) =R^{P,P}\left( N,K,c,E\right) .$

$\left( c\right) $ $U\left( N,M,t\right) =R^{CEA,P}\left( N,K,c,E\right) .$
\end{theorem}


The proof of Theorem \ref{claims results} can be found in the appendix.

From Theorem $\ref{claims results} \left( a\right)$ we obtain that the
pro-rata index and the user-centric index can be rationalized as weighted
proportional rules. We have seen in Theorem \ref{char family} that they are
weighted indices. Thus, one might naturally conjecture whether there is a
bijection between weighted proportional rules (for claims problems) and
weighted indices (for streaming problems). The answer is not. We can indeed
find weighted proportional rules that can not be obtained through a weighted
index and weighted indices such that the induced rule to allocate awards in
streaming problems is not a weighted proportional rule.

Theorem $\ref{claims results} \left( b\right) $ and $\ref{claims results}
\left( c\right) $ allow us to consider both indices from another
perspective. As the statements indicate, the allcation rules they induce
(for streaming problems) can actually be described as two-stage (bankruptcy)
rules where we first decide the importance of each user an then the
importance of each artist for each user, which is computed as the sum over
all users. The pro-rata and user-centric indices measure the importance of
each artist in different ways. For the latter, all users have the same
importance, whereas for the former the importance of each user is
proportional to the user's streams. They then measure the importance of each
artist for each user in the same way; namely, proportionally to the artists'
streams.


To conclude with this section, we reiterate that Theorem \ref{claims results}
states that both indices could be rationalized as a combination of two
well-known rules from the literature on bankruptcy problems. Besides, both
of them can also be seen as members of the same family of weighted
proportional rules. Thus, we can conclude that (in contrast with the
previous sections) the analysis in this section does not favor one of the
indices over the other.

\section{Conclusion}

We have analyzed in this paper the problem of sharing the revenues raised
from subscription fees to music platforms among participating artists. Our
analysis has highlighted two central methods (pro-rata and user-centric)
which can actually be seen as focal (and somewhat polar) members of a family
of methods which evaluate artists by the weighted aggregation of users'
streaming choices. The weight assigned to each user might actually depend on
the user herself and her whole streaming profile. We therefore provide a
solid common ground for both methods, in the form of the characterization
result for the whole family. We, nevertheless, provide additional
(normative, as well as game-theoretical) arguments to favor the user-centric
method with respect to the pro-rata method. To wit, we show that the former
satisfies two natural and appealing axioms (reasonable lower bound and
click-fraud proofness) that the second violates. They are somewhat connected
to a feature the second exhibits (whereas the first does not); namely,
cross-subsidization between high- and low-streaming-volume users.
Furthermore, the former satisfies core-selection, while the latter does not
(which implies that it does not guarantee allocations preventing incentives
for artist to leave the platform) 


To complement the above analysis, we simply stress that two aspects are
relevant when it comes to measuring the importance of artists (within a
platform). On the one hand, the users playing content from the artist. On
the other hand, the streaming times the artist achieved. Although both
aspects are related, they may differ largely. For instance, two artists may
have been listened by the same number of users but with very different
streaming times. We believe the pro-rata index manages well the streaming
times but not the number of users. 
The user-centric index manages well both the streaming times and the number
of users.

Our analysis should contribute to the debate between the pro-rata and
user-centric methods in the music industry. Nevertheless, the discussion in
that industry nowadays goes beyond the debate between those two methods. For
instance, the French streaming service Deezer claims to be pioneer in fair
payments to artist, ``being a main advocate for a re-evaluation of music
streaming's economic model". 
In March 2023, Deezer announced an initiative with Universal Music Group,
the world leader in music-based entertainment, to explore new streaming
models that better align the interests of artists, fans and streaming
services. Using deep data analysis, this partnership aims to improve the
fairness of the current streaming model in various ways, whether by helping
artists monetize their music better or by eliminating issues within the
current system. This initiative will not prioritize just the most-streamed
artists on the platform, but will level the playing field for artists at
every stage of their career and benefit the wider music community as a whole.%
\footnote{%
See https://www.deezer-blog.com/how-much-does-deezer-pay-artists/} We
believe that some of these goals might be achieved with other members of the
more general family of weigthed indices we characterize in this paper. As we
mentioned above, weigthed indices are precisely constructed on the premise
that each artist is assigned the weighted aggregation of streamings, accross
users, with the weight depending on the user and her streaming profile. 

\section{Appendix}


\subsection*{Proof of Theorem \protect\ref{char family}}

$\left( a\right) $ We first prove that each weighted index satisfies the two
axioms.

For each $\omega,$ $I^{\omega}$ satisfies \textit{homogeneity}. Let $\left(
N,M,t\right)\in \mathcal{P} $, $i,i^{\prime }\in N,$ and $\lambda \in 
\mathbb{N}_{+}$ such that $t_{ij}=\lambda t_{i^{\prime }j}$ for all $j\in M,$
$\ $ 
\begin{equation*}
I_{i}^{\omega}\left( N,M,t\right) =\sum_{j\in M}\omega\left( j,t._{j}\right)
t_{ij}=\sum_{j\in M}\omega\left( j,t._{j}\right) \lambda t_{i^{\prime }j}
=\lambda I_{i^{\prime }}^{\omega}\left( N,M,t\right) .
\end{equation*}

For each $\omega ,$ $I^{\omega }$ satisfies \textit{additivity}. Let $\left(
N,M^{1},t^{1}\right) ,\left( N,M^{2},t^{2}\right) ,\left( N,M,t\right) \in 
\mathcal{P}$ such that $M=M^{1}\cup M^{2}$, $t_{ij}=t_{ij}^{1}$ when $j\in
M^{1}$ and $t_{ij}=t_{ij}^{2}$ when $j\in M^{2}.$ For each $i\in N$, 
\begin{eqnarray*}
I_{i}^{\omega }\left( N,M^{1},t^{1}\right) +I_{i}^{\omega }\left(
N,M^{2},t^{2}\right) &=&\sum_{j\in M^{1}}\omega \left( j,t_{.j}^{1}\right)
t_{ij}^{1}+\sum_{j\in M^{2}}\omega \left( j,t_{.j}^{2}\right) t_{ij}^{2} \\
&=&\sum_{j\in M}\omega \left( j,t_{.j}\right) t_{ij}=I_{i}^{\omega }\left(
N,M,t\right) .
\end{eqnarray*}

Conversely, we now prove that if an index $I$ satisfies the two axioms, then
there exists a weight system for streams $\omega$ such that $I=I^{\omega}.$

Let $\left( N,M,t\right)\in \mathcal{P} $. By \textit{additivity}, for each $%
i\in N,$ 
\begin{equation*}
I_{i}\left( N,M,t\right) =\sum_{j\in M}I_{i}\left( N,\left\{ j\right\}
,t_{.j}\right) .
\end{equation*}

Let $j\in M.$ By \textit{homogeneity}, for each $i\in N$ such that $t_{ij}=0$%
, $I_{i}\left( N,\left\{ j\right\} ,t_{.j}\right) =0.$

Recall the basic assumption in our model that $L^{j}\left( N,\left\{
j\right\} ,t_{.j}\right) \neq \varnothing$. Then, let $i\in L^{j}\left(
N,\left\{ j\right\} ,t_{.j}\right) .$ By \textit{homogeneity}, for each $%
i^{\prime }\in $ $L^{j}\left( N,\left\{ j\right\} ,t_{.j}\right) ,$ 
\begin{equation*}
I_{i^{\prime }}\left( N,\left\{ j\right\} ,t_{.j}\right) =\frac{t_{i^{\prime
}j}}{t_{ij}}I_{i}\left( N,\left\{ j\right\} ,t_{.j}\right) .
\end{equation*}

Then, 
\begin{eqnarray*}
\sum_{i^{\prime }\in N}I_{i^{\prime }}\left( N,\left\{ j\right\}
,t_{.j}\right) &=&\sum_{i^{\prime }\in L^{j}\left( N,\left\{ j\right\}
,t\right) }I_{i^{\prime }}\left( N,\left\{ j\right\} ,t_{.j}\right) \\
&=&\sum_{i^{\prime }\in L^{j}\left( N,\left\{ j\right\} ,t\right) }\frac{%
t_{i^{\prime }j}}{t_{ij}}I_{i}\left( N,\left\{ j\right\} ,t_{.j}\right) \\
&=&\frac{T^{j}\left( N,\left\{ j\right\} ,t_{.j}\right) }{t_{ij}}I_{i}\left(
N,\left\{ j\right\} ,t_{.j}\right) .
\end{eqnarray*}

Hence, 
\begin{equation*}
I_{i}\left( N,\left\{ j\right\} ,t_{.j}\right) =\frac{\sum\limits_{i^{\prime
}\in N}I_{i^{\prime }}\left( N,\left\{ j\right\} ,t_{.j}\right) }{%
T^{j}\left( N,\left\{ j\right\} ,t_{.j}\right) }t_{ij}.
\end{equation*}

Taking $\omega \left( j,t_{.j}\right) =\frac{\sum\limits_{i^{\prime }\in
N}I_{i^{\prime }}\left( N,\left\{ j\right\} ,t_{.j}\right) }{T^{j}\left(
N,\left\{ j\right\} ,t_{.j}\right) }$ we have that 
\begin{equation*}
I_{i}\left( N,M,t\right) =\sum_{j\in M}I_{i}\left( N,\left\{ j\right\}
,t_{.j}\right) =\sum_{j\in M}\omega \left( j,t_{.j}\right)
t_{ij}=I_{i}^{\omega }\left( N,M,t\right) .
\end{equation*}

\bigskip

$\left( b\right) $ As the pro-rata index is a weighted index, we know it
satisfies \textit{homogeneity} and \textit{additivity}. As for \textit{equal
individual impact of similar users}, let $\left( N,M,t\right) \in \mathcal{P}
$, $i\in N$ and $j,j^{\prime }\in M$ such that $t_{ij}=t_{ij^{\prime }}$.
Then, 
\begin{eqnarray*}
P_{i}\left( N,M\backslash \left\{ j\right\} ,t^{-j}\right) &=&T_{i}\left(
N,M,t^{-j}\right) -t_{ij} \\
&=&T_{i}\left( N,M,t^{-j^{\prime }}\right) -t_{ij^{\prime }} \\
&=&P_{i}\left( N,M\backslash \left\{ j^{\prime }\right\} ,t^{-j^{\prime
}}\right) .
\end{eqnarray*}

Let now $I$ be an index satisfying the three axioms in the second statement.
By the proof of the previous statement, we know that $I=I^{\omega}$ for some
weight system $\omega.$

We now prove that $\omega \left( j,t_{.j}\right) $ does not depend on $j$
and $t_{.j}.$ Let $j\in M$ and $i\in L^{j}\left( N,\left\{ j\right\}
,t_{.j}\right) .$ For each $x>0$ consider the problem $\left( N,\left\{
j^{\prime }\right\} ,t^{x}\right) $ such that $t_{ij^{\prime }}=x$ for all $%
i\in N.$ By \textit{homogeneity}, for each pair $i,i^{\prime }\in N,$ $%
I_{i}\left( N,\left\{ j^{\prime }\right\} ,t^{x}\right) =I_{i^{\prime
}}\left( N,\left\{ j^{\prime }\right\} ,t^{x}\right) .$ Let $\left(
N,M^{\prime },t^{\prime }\right) \in \mathcal{P}$ be such that $M^{\prime
}=\left\{ j,j^{\prime }\right\} $ and for all $i^{\prime }\in N,$ $%
t_{i^{\prime }j}^{\prime }=t_{i^{\prime }j}$ and $t_{i^{\prime }j^{\prime
}}^{\prime }=t_{i^{\prime }j^{\prime }}^{t_{ij}}.$

By \textit{equal individual impact of similar users}, 
\begin{equation*}
I_{i}\left( N,M^{\prime }\backslash \left\{ j\right\} ,t^{-j}\right)
=I_{i}\left( N,M^{\prime }\backslash \left\{ j^{\prime }\right\}
,t^{-j^{\prime }}\right) .
\end{equation*}

Then, 
\begin{eqnarray*}
\omega\left( j,t_{.j}\right) t_{ij} &=&I_{i}\left( N,\left\{ j\right\}
,t_{.j}\right) =I_{i}\left( N,M^{\prime }\backslash \left\{ j^{\prime
}\right\} ,t^{-j^{\prime }}\right) \\
&=&I_{i}\left( N,M^{\prime }\backslash \left\{ j\right\} ,t^{-j}\right)
=I_{i}\left( N,\left\{ j^{\prime }\right\} ,t^{t_{ij}}\right) \\
&=&\omega\left( j^{\prime },t^{t_{ij}}\right) t_{ij}.
\end{eqnarray*}

Hence, $\omega\left( j,t_{.j}\right) =\omega\left( j^{\prime
},t^{t_{ij}}\right) .$

Let $t^{\ast }\in \mathbb{N}_{+}^{N}$ be such that $t_{ij}^{\ast }=t_{ij}$
and $t_{i^{\prime }j}^{\ast }=1$ when $i^{\prime }\neq j.$ Similarly to $%
\omega\left( j,t_{.j}\right) $, we can argue that $\omega\left( j,t^{\ast
}\right) =\omega\left( j^{\prime },t^{t_{ij}}\right) .$

Take $\left( N,\left\{ j\right\} ,t^{\ast }\right) \in \mathcal{P}$ and $%
i^{\prime }\neq i.$ Similarly to $\omega \left( j,t_{.j}\right) $ with $%
i^{\prime }$ instead of $i$ we can argue that $\omega \left( j,t^{\ast
}\right) =\omega \left( j^{\prime },t^{1}\right) .$ Thus, for each $%
j,j^{\prime }\in M$ and each $t_{.j}$ we have that $\omega \left(
j,t_{.j}\right) =\omega \left( j^{\prime },t^{1}\right) .$ Then, $\omega
\left( j,t_{.j}\right) $ does not depend on $j$ and $t_{.j}$ and we can
define 
\begin{equation*}
\lambda =\omega \left( j,t_{.j}\right) .
\end{equation*}

Now, for each $\left( N,M,t\right) $ and each $i\in N,$ 
\begin{equation*}
I_{i}\left( N,M,t\right) =I_{i}^{\omega}\left( N,M,t\right) =\sum_{j\in
M}\lambda t_{ij}=\lambda T_{i}=\lambda P_{i}\left( N,M,t\right) .
\end{equation*}

Then, $I$ is a (positive) linear transformation of $P$ and hence it
coincides with $P$ (recall that we define indices up to a positive linear
transformation, 
because the resulting ones generate the same allocation of rewards among
artists).

\bigskip

$\left( 3\right) $ As the user-centric index is a weighted index, we know it
satisfies \textit{homogeneity} and \textit{additivity}. As for \textit{equal
global impact of users}, let $\left( N,M,t\right) \in \mathcal{P}$ and $j\in
M,$ 
\begin{equation*}
\sum_{i\in N}U_{i}\left( N,M\backslash \left\{ j\right\} ,t^{-j}\right)
=\sum_{i\in N}\sum_{k\in M\backslash \left\{ j\right\} }\frac{t_{ik}}{T^{k}}%
=\sum_{k\in M\backslash \left\{ j\right\} }\sum_{i\in N}\frac{t_{ik}}{T^{k}}%
=\sum_{k\in M\backslash \left\{ j\right\} }1=m-1.
\end{equation*}

Thus, $\sum\limits_{i\in N}U_{i}\left( N,M\backslash \left\{ j\right\}
,t^{-j}\right) $ does not depend on $j\in M$ and hence the user-centric
index satisfies \textit{equal global impact of users}.

Let now $I$ be an index satisfying the axioms in the third statement. By the
proof of the first statement, 
for each $\left( N,M,t\right) \in \mathcal{P}$ and each $i\in N,$ 
\begin{equation*}
I_{i}\left( N,M,t\right) =\sum_{j\in M}\omega \left( j,t_{.j}\right) t_{ij},
\end{equation*}%
where%
\begin{equation*}
\omega \left( j,t_{.j}\right) =\frac{\sum\limits_{i^{\prime }\in
N}I_{i^{\prime }}\left( N,\left\{ j\right\} ,t_{.j}\right) }{T^{j}\left(
N,\left\{ j\right\} ,t_{.j}\right) }
\end{equation*}

Let $j,j^{\prime }\in M$ and $M^{\ast }=\left\{ j,j^{\prime }\right\} .$ By 
\textit{equal global impact of users}, 
\begin{equation*}
\sum_{i\in N}I_{i}\left( N,M^{\ast }\backslash \left\{ j\right\}
,t_{.j^{\prime }}\right) =\sum_{i\in N}I_{i}\left( N,M^{\ast }\backslash
\left\{ j^{\prime }\right\} ,t_{.j}\right) .
\end{equation*}

Thus, 
\begin{equation*}
\sum_{i\in N}I_{i}\left( N,\left\{ j^{\prime }\right\} ,t_{.j^{\prime
}}\right) =\sum_{i\in N}I_{i}\left( N,\left\{ j\right\} ,t_{.j}\right) .
\end{equation*}

Then, $\sum\limits_{i\in N}I_{i}\left( N,\left\{ j\right\} ,t_{.j}\right) $
does not depend on $j$ and $t_{.j}$ and we can define 
\begin{equation*}
\lambda =\sum_{i\in N}I_{i}\left( N,\left\{ j\right\} ,t_{.j}\right) .
\end{equation*}

As $T^{j}\left( N,\left\{ j\right\} ,t_{.j}\right) =T^{j}\left( N,M,t\right)$%
, it follows that 
\begin{equation*}
I_{i}\left( N,M,t\right) =\sum_{j\in M}\frac{\lambda }{T^{j}\left(
N,M,t\right) }t_{ij}=\lambda U_{i}\left( N,M,t\right) .
\end{equation*}

Then, $I$ is a (positive) linear transformation of $U$ and hence it
coincides with $U$.



\subsection*{Independence of the axioms of Theorem \protect\ref{char family}}

We now show that all the axioms used in the previous characterizations are
independent.

$\left( a\right) $ Let $I^{1}$ be the uniform index that assigns to each
artist a constant number. Namely, for each $\left( N,M,t\right) $ and each $%
i\in N$, $I_{i}^{1}\left( N,M,t\right) =1$. 
$I^{1}$ satisfies \textit{additivity} but not \textit{homogeneity}.

Let $I^{2}$ be defined as follows. For each $\left( N,M,t\right) $ and each $%
i\in N,$ 
\begin{equation*}
I_{i}^{2}\left( N,M,t\right) =\sum_{j=1}^{m}\frac{t_{ij}+T_{i}}{%
T^{j}+\sum\limits_{i^{\prime }\in N}T_{i^{\prime }}}.
\end{equation*}%
$I^{2}$ satisfies \textit{homogeneity} but not \textit{additivity}.


$\left( b\right) $ Let $I^{3}$ be defined as follows. For each $\left(
N,M,t\right) $ and each $i\in N,$ 
\begin{equation*}
I_{i}^{3}\left( N,M,t\right) =\sum_{j=1}^{m}t_{ij}^{2}.
\end{equation*}%
$I^{3}$ satisfies \textit{additivity} and \textit{equal individual impact of
similar users} but not \textit{homogeneity}.

$I^{2}$, defined above, satisfies \textit{homogeneity} and \textit{equal
individual impact of similar users}, but not \textit{additivity}.

The user-centric index satisfies \textit{homogeneity} and \textit{additivity}
but not \textit{equal individual impact of similar users}.

$\left( c\right) $ $I^{1}$, defined above, satisfies \textit{additivity} and 
\textit{equal global impact of users}, but not \textit{homogeneity}.

Let $I^{4}$ be defined as follows. For each $\left( N,M,t\right) $ and each $%
i\in N,$ 
\begin{equation*}
I_{i}^{4}\left( N,M,t\right) =\frac{T_{i}}{\sum\limits_{i^{\prime }\in
N}T_{i^{\prime }}}m.
\end{equation*}%
$I^{4}$ satisfies \textit{homogeneity} and \textit{equal global impact of
users}, but not \textit{additivity}.

The pro-rata index satisfies \textit{homogeneity} and \textit{additivity},
but not \textit{equal global impact of users}.





\subsection*{Reasonable lower bound and click-fraud-proofness}


The user-centric index satisfies reasonable lower bound. Let $\left(
N,M,t\right) \in \mathcal{P}$, $C\subset M$ and $L^{C}=\bigcup\limits_{j\in
C}L^{j}.$ Then, 
\begin{eqnarray*}
\sum_{i\in L^{C}}\frac{U_{i}\left( N,M,t\right) }{\sum\limits_{i^{\prime
}\in N}U_{i^{\prime }}\left( N,M,t\right) }m &=&\sum_{i\in L^{C}}U_{i}\left(
N,M,t\right) =\sum_{i\in L^{C}}\sum_{j\in M}\frac{t_{ij}}{T^{j}} \\
&\geq &\sum_{i\in L^{C}}\sum_{j\in C}\frac{t_{ij}}{T^{j}}=\sum_{j\in
C}\sum_{i\in L^{C}}\frac{t_{ij}}{T^{j}} \\
&=&\sum_{j\in C}1=\left\vert C\right\vert .
\end{eqnarray*}

The pro-rata index does not satisfy it. To show that, consider Example \ref%
{ex 2,2,a} and let $C=\left\{ a\right\} .$ Then $L^{a}=\left\{ 1\right\} $
but 
\begin{equation*}
\frac{P_{1}\left( N,M,t\right) }{\sum\limits_{i^{\prime }\in N}P_{i^{\prime
}}\left( N,M,t\right) }2=\frac{20}{100}<1.
\end{equation*}

\bigskip

The user-centric index satisfies click-fraud-proofness. To show this, let $%
\left( N,M,t\right) $, $\left( N,M,t^{\prime }\right) $, and $j\in M$ such
that $t_{ij^{\prime }}=t_{ij^{\prime }}^{\prime }$ for all $j^{\prime }\in
M\backslash \left\{ j\right\} .$ Then, for all $i\in N,$ 
\begin{eqnarray*}
\left\vert \frac{U_{i}\left( N,M,t\right) }{\sum\limits_{i^{\prime }\in
N}U_{i^{\prime }}\left( N,M,t\right) }m-\frac{U_{i}\left( N,M,t^{\prime
}\right) }{\sum\limits_{i^{\prime }\in N}U_{i^{\prime }}\left( N,M,t^{\prime
}\right) }m\right\vert &=&\left\vert U_{i}\left( N,M,t\right) -U_{i}\left(
N,M,t^{\prime }\right) \right\vert \\
&=&\left\vert \frac{t_{ij}}{T^{j}\left( N,M,t\right) }-\frac{t_{ij}^{\prime }%
}{T^{j}\left( N,M,t^{\prime }\right) }\right\vert \\
&\leq &1,
\end{eqnarray*}

where the last inequality is true because $\frac{t_{ij}}{T^{j}\left(
N,M,t\right) },\frac{t_{ij}^{\prime }}{T^{j}\left( N,M,t^{\prime }\right) }%
\in \left[ 0,1\right] $.

The pro-rata index, however, does not satisfy it. To show that, consider
Example \ref{ex 2,2,a}. If we let 
\begin{equation*}
t^{\prime }=\left( 
\begin{array}{cc}
10 & 0 \\ 
0 & 2%
\end{array}%
\right) ,
\end{equation*}%
we obtain%
\begin{equation*}
\left\vert \frac{P_{2}\left( N,M,t\right) }{\sum\limits_{i^{\prime }\in
N}P_{i^{\prime }}\left( N,M,t\right) }m-\frac{P_{2}\left( N,M^{^{\prime
}},t^{\prime }\right) }{\sum\limits_{i^{\prime }\in N}P_{i^{\prime }}\left(
N,M^{\prime },t^{\prime }\right) }m\right\vert =\left\vert \frac{90}{100}2-%
\frac{2}{12}2\right\vert =\frac{22}{15}>1.
\end{equation*}

\bigskip

\subsection*{Proof of Theorem \protect\ref{core all}}

We prove Theorem \ref{core all} using a classical result from the literature
on cooperative game theory. We first introduce some notation. We then
explain the classical result. Finally, we apply it to our game $\left(
N,v\right) .$

The unanimity game $u_{R}$ associated with the nonempty coalition $R\subset
N $ is defined as $u_{R}\left( S\right) =1$ when $R\subset S$ and $%
u_{R}\left( S\right) =0$ otherwise.

Harsanyi (1959) proved that for every $TU$ game $\left( N,v\right) $, there
exist unique weights $\Delta _{v}\left( R\right) \in \mathbb{R}$ such that $%
v=\sum\limits_{\varnothing \neq R\subset N}\Delta _{v}\left( R\right) u_{R}.$
The weights $\Delta _{v}\left( R\right) $ are called the (Harsanyi)
dividends of $v.$

Given a $TU$ game $v,$ the Harsanyi set $H(N,v)$ (Vasil'ev 1978, 1981) is
defined as the set of allocations obtained by distributing the dividend of
any coalition $R$ in any way among the agents in $R.$ We now introduce it
formally.\footnote{%
The presentation we consider here follows the one in van den Brink et al.,
(2014).}

We say that $p=\left( p^{R}\right) _{\varnothing \neq R\subset N}$ is a
sharing system if for each $\varnothing \neq R\subset N$ we have that $%
p^{R}\in \mathbb{R}^{N},$ $p_{i}^{R}\geq 0$ for all $i\in R,$ $p_{i}=0$ for
all $i\in N\backslash R,$ and $\sum\limits_{i\in N}p_{i}^{R}=1.$ Each
sharing system $p$ induces the allocation $y^{p}$ where for each $i\in N$, 
\begin{equation*}
y_{i}^{p}=\sum_{i\in R}\Delta _{v}\left( R\right) p_{i}^{R}.
\end{equation*}

The Harsanyi set is defined as 
\begin{equation*}
H\left( N,v\right) =\left\{ y^{p}:p\text{ is a sharing system}\right\} .
\end{equation*}

Vasil'ev (1981) proved that if all dividends of $\left( N,v\right) $ are
nonnegative, $C\left( N,v\right) =H\left( N,v\right) $.

It is straightforward to check that given a streaming problem $\left(
N,M,t\right) $, 
\begin{equation*}
v=\sum_{\varnothing \neq R\subset N}\left\vert j\in M:L^{j}=R\right\vert
u_{R}.
\end{equation*}

Then, for each $\varnothing \neq R\subset N,$ $\Delta _{v}\left( R\right)
=\left\vert j\in M:L^{j}=R\right\vert \geq 0.$ Thus, for our result, it
suffices to prove that $H\left( N,v\right) =A\left( N,M,t\right) .$

We first prove that $H\left( N,v\right) \subset A\left( N,M,t\right) .$ Let $%
p$ be a sharing system. We must prove that $y^{p}\in A\left( N,M,t\right) .$
We define $\left( x^{j}\right) _{j\in M}$ and $x=\sum\limits_{j\in
M}x^{j}\in A\left( N,M,t\right) $ as follows. Given $j\in M$ we take $%
x^{j}=p^{L^{j}}.$ Because of the definition of a sharing system we have that 
$x^{j}$ satisfies the condition of $A\left( N,M,t\right) .$ Besides, for
each $i\in N$ 
\begin{equation*}
x_{i}=\sum_{j\in M}x_{i}^{j}=\sum_{j\in M:i\in
L^{j}}p_{i}^{L^{j}}=\sum_{i\in R\subset N}\left\vert j\in
M:L^{j}=R\right\vert p_{i}^{R}=y_{i}^{p}.
\end{equation*}

We now prove that $A\left( N,M,t\right) \subset H\left( N,v\right) .$ Let $%
\left( x^{j}\right) _{j\in M}$ and $x=\sum\limits_{j\in M}x^{j}\in A\left(
N,M,t\right) .$ We define $p$ as follows. For each $R\subset N$ and $i\in N,$
\begin{equation*}
p_{i}^{R}=\left\{ 
\begin{tabular}{ll}
$\frac{\sum\limits_{j\in M,L^{j}=R}x_{i}^{j}}{\left\vert j\in
M,L^{j}=R\right\vert }$ & $\text{when }\left\vert j\in M,L^{j}=R\right\vert
>0$ \\ 
$\frac{1}{\left\vert N\right\vert }$ & when $\left\vert j\in
M,L^{j}=R\right\vert =0.$%
\end{tabular}%
\right.
\end{equation*}

We prove that $p$ is a sharing system. Let $R\subset N.$ Since $%
x_{i}^{j}\geq 0$ for all $i\in N$ and $j\in M,$ $p_{i}^{R}\geq 0$ when $i\in
R.$ Since $x_{i}^{j}=0$ when $i\in N\backslash L^{j},$ $p_{i}^{R}=0$ when $%
i\in N\backslash R.$ If $\left\vert j\in M,L^{j}=R\right\vert =0,$ obviously 
$\sum\limits_{i\in N}p_{i}^{R}=1.$ Assume that $\left\vert j\in
M,L^{j}=R\right\vert >0.$ Now, 
\begin{equation*}
\sum\limits_{i\in N}p_{i}^{R} =\sum\limits_{i\in N}\frac{\sum\limits_{j\in
M,L^{j}=R}x_{i}^{j}}{\left\vert j\in M,L^{j}=R\right\vert }%
=\sum\limits_{j\in M,L^{j}=R}\frac{\sum\limits_{i\in N}x_{i}^{j}}{\left\vert
j\in M,L^{j}=R\right\vert }=\sum\limits_{j\in M,L^{j}=R}\frac{1}{\left\vert
j\in M,L^{j}=R\right\vert }=1.
\end{equation*}

Besides, for each $i\in N,$ 
\begin{eqnarray*}
y_{i}^{p} &=&\sum_{i\in R\subset N}\left\vert j\in M:L^{j}=R\right\vert
p_{i}^{R} \\
&=&\sum_{i\in R\subset N}\left\vert j\in M:L^{j}=R\right\vert \frac{%
\sum\limits_{j\in M,L^{j}=R}x_{i}^{j}}{\left\vert j\in M,L^{j}=R\right\vert }
\\
&=&\sum_{i\in R\subset N}\sum\limits_{j\in
M,L^{j}=R}x_{i}^{j}=\sum\limits_{j\in M}\sum_{i\in
L_{j}}x_{i}^{j}=\sum\limits_{j\in M}x_{i}^{j} \\
&=&x_{i}.
\end{eqnarray*}%
%
%
%
%
%
%
%
%
%
%




\subsection*{Proof of Theorem \protect\ref{char user-centric}}

We have already seen that the user-centric index satisfies \textit{%
homogeneity}, \textit{additivity}, and \textit{core selection}.

Conversely, let $I$ be an index satisfying the axioms. Let $\left(
N,M,t\right) \in \mathcal{P}^{\ast }$ and $j\in M.$ We consider the problem $%
\left( N,\left\{ j\right\} ,t_{.j}\right) \in \mathcal{P}^{\ast }$. By 
\textit{core selection}, taking $S=$ $L^{j}\left( N,\left\{ j\right\}
,t_{.j}\right) \neq N$, we have 
\begin{equation*}
\sum_{i\in S}\frac{I_{i}\left( N,\left\{ j\right\} ,t_{.j}\right) }{%
\sum\limits_{i^{\prime }\in N}I_{i^{\prime }}\left( N,\left\{ j\right\}
,t_{.j}\right) }\geq v_{\left( N,\left\{ j\right\} ,t_{.j}\right) }\left(
S\right) =1,
\end{equation*}%
which implies 
\begin{equation*}
\sum_{i\in L^{j}\left( N,\left\{ j\right\} ,x\right) }I_{i}\left( N,\left\{
j\right\} ,t_{.j}\right) \geq \sum\limits_{i^{\prime }\in N}I_{i^{\prime
}}\left( N,\left\{ j\right\} ,t_{.j}\right) .
\end{equation*}%
Thus, 
\begin{equation*}
\sum_{i\in N\backslash L^{j}\left( N,\left\{ j\right\} ,x\right)
}I_{i}\left( N,\left\{ j\right\} ,t_{.j}\right) \leq 0.
\end{equation*}%
And, therefore, 
\begin{equation}
I_{i}\left( N,\left\{ j\right\} ,t_{.j}\right) =0\text{ for each }i\in
N\backslash L^{j}\left( N,\left\{ j\right\} ,t_{.j}\right) .
\label{0 en N - Lj}
\end{equation}

Now, let $i,i^{\prime }\in L^{j}\left( N,\left\{ j\right\} ,t_{.j}\right) .$
By \textit{homogeneity}, 
\begin{equation*}
I_{i^{\prime }}\left( N,\left\{ j\right\} ,t_{.j}\right) =\frac{t_{i^{\prime
}j}}{t_{ij}}I_{i}\left( N,\left\{ j\right\} ,t_{.j}\right) .
\end{equation*}

Then, for each $i\in L^{j}\left( N,\left\{ j\right\} ,t_{.j}\right) $,%
\begin{equation*}
\sum_{i^{\prime }\in N}I_{i^{\prime }}\left( N,\left\{ j\right\}
,t_{.j}\right) =\frac{\sum\limits_{i^{\prime }\in N}t_{i^{\prime }j}}{t_{ij}}%
I_{i}\left( N,\left\{ j\right\} ,t_{.j}\right) .
\end{equation*}

Hence, 
\begin{equation*}
I_{i}\left( N,\left\{ j\right\} ,t_{.j}\right) =\left[ \sum_{i^{\prime }\in
N}I_{i^{\prime }}\left( N,\left\{ j\right\} ,t_{.j}\right) \right] \frac{%
t_{ij}}{\sum\limits_{i^{\prime }\in N}t_{i^{\prime }j}}=\left[
\sum_{i^{\prime }\in N}I_{i^{\prime }}\left( N,\left\{ j\right\}
,t_{.j}\right) \right] U_{i}\left( N,\left\{ j\right\} ,t_{.j}\right) .
\end{equation*}

We now prove the following claim.

\textbf{Claim}. For each pair $j^{1},j^{2}\in M$, 
such that $j^{1}\neq j^{2}$ and $L^{j^{1}}\left( N,\left\{ j^{1}\right\}
,t_{.j^{1}}\right) \cap L^{j^{2}}\left( N,\left\{ j^{2}\right\}
,t_{.j^{2}}\right) =\varnothing $ we have 
\begin{equation*}
\sum_{i^{\prime }\in N}I_{i^{\prime }}\left( N,\left\{ j^{1}\right\}
,t_{.j^{1}}\right) =\sum_{i^{\prime }\in N}I_{i^{\prime }}\left( N,\left\{
j^{2}\right\} ,t_{.j^{2}}\right) .
\end{equation*}

Suppose that the claim is false. Then, without loss of generality, assume
that 
\begin{equation*}
\sum_{i^{\prime }\in N}I_{i^{\prime }}\left( N,\left\{ j^{1}\right\}
,t_{.j^{1}}\right) <\sum_{i^{\prime }\in N}I_{i^{\prime }}\left( N,\left\{
j^{2}\right\} ,t_{.j^{2}}\right) .
\end{equation*}

Let $\left( N,\left\{ j^{1},j^{2}\right\} ,\hat{t}\right) \in \mathcal{P}%
^{\ast }$ where $\hat{t}$ denotes the restriction of $t$ to $\left\{
j^{1},j^{2}\right\} .$ By \textit{core selection}, taking $S=$ $%
L^{j^{1}}\left( N,\left\{ j^{1}\right\} ,t_{.j^{1}}\right) ,$ 
\begin{equation*}
1=v_{\left( N,\left\{ j^{1},j^{2}\right\} ,\hat{t}\right) }\left( S\right)
\leq \sum_{i\in S}\frac{I_{i}\left( N,\left\{ j^{1},j^{2}\right\} ,\hat{t}%
\right) }{\sum\limits_{i^{\prime }\in N}I_{i^{\prime }}\left( N,\left\{
j^{1},j^{2}\right\} ,\hat{t}\right) }2
\end{equation*}

By \textit{additivity}$,$ 
\begin{eqnarray*}
\sum\limits_{i^{\prime }\in N}I_{i^{\prime }}\left( N,\left\{
j^{1},j^{2}\right\} ,\hat{t}\right) &=&\sum_{i^{\prime }\in N}I_{i^{\prime
}}\left( N,\left\{ j^{1}\right\} ,t_{.j^{1}}\right) +\sum_{i^{\prime }\in
N}I_{i^{\prime }}\left( N,\left\{ j^{2}\right\} ,t_{.j^{2}}\right) \text{
and } \\
\sum_{i\in L^{j^{1}}\left( N,\left\{ j^{1}\right\} ,t_{.j^{1}}\right)
}I_{i}\left( N,\left\{j^{1},j^{2}\right\} ,\hat{t}\right) &=&\sum_{i\in
L^{j^{1}}\left( N,\left\{ j^{1}\right\} ,t_{.j^{1}}\right) }I_{i}\left(
N,\left\{ j^{1}\right\} ,t_{.j^{1}}\right) +\sum_{i\in L^{j^{1}}\left(
N,\left\{ j^{1}\right\} ,t_{.j^{1}}\right) }I_{i}\left( N,\left\{
j^{2}\right\} ,t_{.j^{2}}\right) .
\end{eqnarray*}

By $\left( \ref{0 en N - Lj}\right) ,$ 
\begin{eqnarray*}
\sum_{i\in L^{j^{1}}\left( N,\left\{ j^{1}\right\} ,t_{.j^{1}}\right)
}I_{i}\left( N,\left\{ j^{1}\right\} ,t_{.j^{1}}\right) &=&\sum_{i\in
N}I_{i}\left( N,\left\{ j^{1}\right\} ,t_{.j^{1}}\right) \text{ and } \\
\sum_{i\in L^{j^{1}}\left( N,\left\{ j^{1}\right\} ,t_{.j^{1}}\right)
}I_{i}\left( N,\left\{ j^{2}\right\} ,t_{.j^{2}}\right) &=&0.
\end{eqnarray*}

Then, 
\begin{eqnarray*}
1 &\leq &\frac{\sum\limits_{i\in N}I_{i}\left( N,\left\{ j^{1}\right\}
,t_{.j^{1}}\right) }{\sum\limits_{i^{\prime }\in N}I_{i^{\prime }}\left(
N,\left\{ j^{1}\right\} ,t_{.j^{1}}\right) +\sum\limits_{i^{\prime }\in
N}I_{i^{\prime }}\left( N,\left\{ j^{2}\right\} ,t_{.j^{2}}\right) }2 \\
&<&\frac{\sum\limits_{i\in N}I_{i}\left( N,\left\{ j^{1}\right\}
,t_{.j^{1}}\right) }{\sum\limits_{i^{\prime }\in N}I_{i^{\prime }}\left(
N,\left\{ j^{1}\right\} ,t_{.j^{1}}\right) +\sum\limits_{i^{\prime }\in
N}I_{i^{\prime }}\left( N,\left\{ j^{1}\right\} ,t_{.j^{1}}\right) }2=1,
\end{eqnarray*}%
which is a contradiction. Hence, the claim holds.

We now prove that, for each pair $j^{1},j^{2}\in M$, with $j^{1}\neq j^{2}$ 
\begin{equation*}
\sum_{i^{\prime }\in N}I_{i^{\prime }}\left( N,\left\{ j^{1}\right\}
,t_{.j^{1}}\right) =\sum_{i^{\prime }\in N}I_{i^{\prime }}\left( N,\left\{
j^{2}\right\} ,t_{.j^{2}}\right) .
\end{equation*}

By the claim, the result holds when $L^{j^{1}}\left( N,\left\{ j^{1}\right\}
,t_{.j^{1}}\right) \cap L^{j^{2}}\left( N,\left\{ j^{2}\right\}
,t_{.j^{2}}\right) =\varnothing .$ Assume then otherwise. 
We consider two cases.

\begin{enumerate}
\item $L^{j^{1}}\left( N,\left\{ j^{1}\right\} ,t_{.j^{1}}\right) \supset
L^{j^{2}}\left( N,\left\{ j^{2}\right\} ,t_{.j^{2}}\right) $.\footnote{%
The case $L^{j^{1}}\left( N,\left\{ j^{1}\right\} ,t_{.j^{1}}\right) \subset
L^{j^{2}}\left( N,\left\{ j^{2}\right\} ,t_{.j^{2}}\right) $ is similar and,
thus, we omit it.}

As $\left( N,\left\{ j^{1}\right\} ,t_{.j^{1}}\right) \in \mathcal{P}^{\ast }
$, it follows that $S=N\setminus L^{j^{1}}\left( N,\left\{ j^{1}\right\}
,t_{.j^{1}}\right) \neq \varnothing .$ Let $j^{3}\in M\backslash \left\{
j^{1},j^{2}\right\} $ and $t_{.j^{3}}^{S}$ be such that 
\begin{equation*}
t_{ij^{3}}^{S}=\left\{ 
\begin{tabular}{ll}
$1$ & for each $i\in S$, \\ 
$0$ & for each $i\in N\setminus S$.%
\end{tabular}
\right.
\end{equation*}

Then, by the Claim above, 
\begin{equation*}
\sum_{i^{\prime }\in N}I_{i^{\prime }}\left( N,\left\{ j^{1}\right\}
,t_{.j^{1}}\right) =\sum_{i^{\prime }\in N}I_{i^{\prime }}\left( N,\left\{
j^{3}\right\} ,t_{.j^{3}}^{S}\right) =\sum_{i^{\prime }\in N}I_{i^{\prime
}}\left( N,\left\{ j^{2}\right\} ,t_{.j^{2}}\right) .
\end{equation*}

\item $S=L^{j^{1}}\left( N,\left\{ j^{1}\right\} ,t_{.j^{1}}\right)
\setminus L^{j^{2}}\left( N,\left\{ j^{2}\right\} ,t_{.j^{2}}\right) \neq
\varnothing $ and $S^{\prime }=L^{j^{2}}\left( N,\left\{ j^{2}\right\}
,t_{.j^{2}}\right) \setminus L^{j^{1}}\left( N,\left\{ j^{1}\right\}
,t_{.j^{1}}\right) \neq \varnothing .$ Let $t_{.j^{1}}^{S}$ and $%
t_{.j^{2}}^{S^{\prime }}$ be defined as in Case 1. Then, by the Claim above, 
\begin{equation*}
\sum_{i^{\prime }\in N}I_{i^{\prime }}\left( N,\left\{ j^{1}\right\}
,t_{.j^{1}}\right) =\sum_{i^{\prime }\in N}I_{i^{\prime }}\left( N,\left\{
j^{2}\right\} ,t_{.j^{2}}^{S^{\prime }}\right) =\sum_{i^{\prime }\in
N}I_{i^{\prime }}\left( N,\left\{ j^{1}\right\} ,t_{.j^{1}}^{S}\right)
=\sum_{i^{\prime }\in N}I_{i^{\prime }}\left( N,\left\{ j^{2}\right\}
,t_{.j^{2}}\right) .
\end{equation*}
\end{enumerate}

\medskip

To conclude the proof, we define $p=\sum\limits_{i^{\prime }\in
N}I_{i^{\prime }}\left( N,\left\{ j\right\} ,t_{.j}\right) .$ Based on the
above, $p$ is well defined because it neither depends on $j$, nor on $t_{.j}.
$ Then, 
\begin{equation*}
I_{i}\left( N,\left\{ j\right\} ,t_{.j}\right) =pU_{i}\left( N,\left\{
j\right\} ,t_{.j}\right) .
\end{equation*}

By \textit{additivity}, 
\begin{equation*}
I_{i}\left( N,M,t\right) =\sum_{j\in M}I_{i}\left( N,\left\{ j\right\}
,t_{.j}\right) =\sum_{j\in M}pU_{i}\left( N,\left\{ j\right\} ,t_{.j}\right)
=pU_{i}\left( N,M,t\right) .
\end{equation*}

Thus, $I$ is also the user-centric index, as desired.


\subsection*{Independence of the axioms of Theorem \protect\ref{char
user-centric}}

We now show that all the axioms used in the previous characterization are
independent.

Let $I^{5}$ be defined as follows. For each $\left( N,M,t\right) $ and each $%
i\in N,$ 
\begin{equation*}
I_{i}^{5}\left( N,M,t\right) =\sum_{j\in M:i\in L^{j}\left( N,\left\{
j\right\} ,t_{.j}\right) }\frac{1}{\left\vert L^{j}\left( N,\left\{
j\right\} ,t_{.j}\right) \right\vert }.
\end{equation*}%
$I^{5}$ satisfies all axioms but \textit{homogeneity}.

$I^{2}$, defined above, satisfies all axioms but \textit{additivity}.

The pro-rata satisfies all axioms but \textit{core selection}.

\bigskip

\subsection*{Proof of Theorem \protect\ref{claims results}}

$\left( a\right) $ We consider the weight function $\omega ^{P}$ given by 
\begin{equation*}
\omega _{j}^{P}\left( \left( C^{j^{\prime }}\right) _{j^{\prime }\in
K},E\right) =\frac{C^{j}}{\sum\limits_{j^{\prime }\in K}C^{j^{\prime }}}.
\end{equation*}

For each $i\in N,$ 
\begin{equation*}
P_{i}^{\omega ^{P}}\left( N,K,c,E\right) =\sum_{j\in M}\frac{t_{ij}}{T^{j}}%
\frac{T^{j}}{\sum\limits_{j^{\prime }\in K}T^{j^{\prime }}}m=\frac{%
\sum\limits_{j\in M}t_{ij}}{\sum\limits_{j^{\prime }\in K}T^{j^{\prime }}}m=%
\frac{T_{i}}{\sum\limits_{i^{\prime }\in N}T_{i^{\prime }}}m=R_{i}^{P}\left(
N,M,t\right) .
\end{equation*}

Thus, $R^{P}$ is the weighted proportional rule associated to $\omega ^{P}$.

We now take the weight function $\omega ^{U}$ given by 
\begin{equation*}
\omega _{j}^{U}\left( \left( C^{j^{\prime }}\right) _{j^{\prime }\in
K},E\right) =\frac{1}{\left\vert K\right\vert }.
\end{equation*}

For each $i\in N,$ 
\begin{equation*}
P_{i}^{\omega ^{U}}\left( N,K,c,E\right) =\sum_{j\in M}\frac{t_{ij}}{T^{j}}%
\frac{1}{m}m=\sum_{j\in M}\frac{t_{ij}}{T^{j}}=U_{i}\left( N,M,t\right) .
\end{equation*}

Thus $U$ is the weighted proportional rule associated to $\omega ^{U}$.

\bigskip

$\left( b\right) $ Given $i\in N,$ we compute $R_{i}^{P,P}\left(
N,K,c,E\right) $

First stage. For each $j\in K,$ 
\begin{equation*}
P_{j}(K,c^{K},E)=\frac{T^{j}}{\sum\limits_{j^{\prime }\in K}T^{j^{\prime }}}%
m.
\end{equation*}

Second stage. 
\begin{equation*}
P_{i}(N,c_{.j},P_{j}\left( K,c^{K},E\right) )=\frac{t_{ij}}{T^{j}}\frac{T^{j}%
}{\sum\limits_{j^{\prime }\in M}T^{j^{\prime }}}m=\frac{t_{ij}}{%
\sum\limits_{j^{\prime }\in M}T^{j^{\prime }}}m.
\end{equation*}

Thus, for each $i\in N$, 
\begin{eqnarray*}
R_{i}^{P,P}(R,N,E,C) &=&\sum_{j\in K}P_{i}(N,c_{.j},P_{j}\left(
K,c^{K},E\right) ) \\
&=&\sum_{j\in M}\frac{t_{ij}}{\sum\limits_{j^{\prime }\in M}T^{j^{\prime }}}%
m=\frac{T_{i}}{\sum\limits_{i^{\prime }\in N}T_{i^{\prime }}}%
m=R_{i}^{P}\left( N,M,t\right) .
\end{eqnarray*}

\bigskip

$\left( c\right) $ Given $i\in N,$ we compute $R_{i}^{CEA,P}\left(
N,K,c,E\right) .$

First stage. For each $j\in K,$ $CEA_{j}(K,c^{K},E)=\min \left\{ \lambda
,T^{j}\right\} $ where $\sum\limits_{j\in K}\min \left\{ \lambda
,T^{j}\right\} =E.$ It is straightforward to check that $\lambda =1.$ Hence, 
\begin{equation*}
CEA_{j}(K,c^{K},E)=1.
\end{equation*}

Second stage. 
\begin{equation*}
P_{i}(N,c_{.j},CEA_{j}(K,c^{K},E))=\frac{t_{ij}}{T^{j}}.
\end{equation*}

Thus, for each $i\in N$, 
\begin{eqnarray*}
R_{i}^{CEA,P}(R,N,E,C) &=&\sum_{j\in K}P_{i}(N,c_{.j},CEA_{j}(K,c^{K},E))) \\
&=&\sum_{j\in M}\frac{c_{ij}}{T^{j}}=U_{i}\left( N,M,t\right) .
\end{eqnarray*}



\bigskip


\bigskip

\bigskip

\end{document}